\documentclass[aps,pre,preprint,groupedaddress]{revtex4}
\usepackage{CJKutf8}

\usepackage{graphicx}
\usepackage{amsmath}
\usepackage{bm}
\usepackage{epsf,epsfig,graphics}
\usepackage{float}
\usepackage{makeidx}
\usepackage{latexsym}
\usepackage{amssymb}
\usepackage{mathrsfs}
\usepackage{amsbsy}
\usepackage{subfigure}

\newcommand{\figws}{7.5cm}
\newcommand{\figwss}{5.cm}

\usepackage{color}
\usepackage{morefloats}
\DeclareMathAlphabet\mathbfcal{OMS}{cmsy}{b}{n}

\begin{document}
\begin{CJK}{UTF8}{gbsn}
\draft
%\twocolumn[\hsize\textwidth\columnwidth\hsize\csname@twocolumnfalse\endcsname

\title{Instantaneous equilibrium Transition for  Brownian systems under time-dependent temperature and potential variations: Reversibility, Heat and work relations, and  Fast Isentropic process}
\author{Yonggun Jun \begin{CJK*}{UTF8}{mj}(전용근)\end{CJK*}\footnote{Email: yonggun@phy.ncu.edu.tw}
and Pik-Yin Lai
\begin{CJK*}{UTF8}{bsmi}(黎璧賢)\end{CJK*}\footnote{Email: pylai@phy.ncu.edu.tw}} 
\affiliation{Department of Physics and Center 
for Complex Systems, National Central University, 300 Zhong-Da Road, Taoyuan City, Taiwan 320}

\hfill\today

\begin{abstract}
The theory of constructing instantaneous equilibrium (ieq) transition under arbitrary time-dependent temperature and potential variation for a Brownian particle is developed. It is shown that it is essential to consider the underdamped dynamics for temperature-changing transitions. The ieq is maintained by a time-dependent auxiliary position and momentum potential, which can be calculated for given time-dependent transition protocols.
Explicit analytic results are derived for the work and heat statistics, energy, and entropy changes for harmonic and non-harmonic trapping potential with arbitrary time-dependent potential parameters and temperature protocols. Numerical solutions of the corresponding Langevin dynamics are computed to confirm the theoretical results. Although ieq transition of the reverse process is not the time-reversal of the ieq transition of the forward process due to the odd-parity of controlling parameters, their phase-space distribution functions restore the time-reversal symmetry, and hence the energy and entropy changes of the ieq of the reverse process are simply the negative of that of the forward process.
Furthermore, it is shown that it is possible to construct an ieq transition that has zero entropy change at a finite transition rate, i.e., a fast ieq isentropic process, and is further demonstrated by explicit Langevin dynamics simulations. Our theory provides fundamental building blocks for designing controlled microscopic heat engine cycles. Implications for constructing an efficient Brownian heat engine are also discussed.
\end{abstract}
%\pacs{ }

\maketitle
\end{CJK}

\section{Introduction}

The Second Law of Thermodynamics, which states that a non-equilibrium process is irreversible with a positive total entropy production, has been a foundation concept in classical thermodynamics for the past two centuries \cite{Callen1985ThermodynamicsThermostatistics}.  
Breakthroughs in theoretical non-equilibrium statistical physics for the last three decades, such as fluctuation theorems \cite{Jarzynski1997,Crooks1999} and stochastic energetics \cite{Sekimoto2010,Seifert2012}, proved to be very successful in a broad range of non-equilibrium processes in small systems in which thermal fluctuations dominate. 
These theories have been  demonstrated  to be quantitatively accurate in various  experimental systems, such as colloidal systems \cite{Carberry2004FluctuationsTrap}, Brownian heat engines \cite{Blickle2011,Martinez2015}, biological systems \cite{Collin2005VerificationEnergies,Liphardt2002EquilibriumEquality}, and quantum systems \cite{Batalhao2014ExperimentalSystem,An2015ExperimentalSystem}. 
Due to the advances of stochastic thermodynamics, we are now at a stage prepared for the applications of stochastic energetics in microscopic thermodynamics, such as manipulating or designing transition paths with special properties. 
One manifested example is the microscopic heat engines of a colloidal particle \cite{Blickle2011,Martinez2015,Albay2021} and a single atom \cite{Rossnagel2016}, which has been intensively studied to understand the efficiency at the maximum power \cite{Schmiedl2008, Esposito2010}, efficiency fluctuations \cite{Verley2014,Polettini2015,Manikandan2019}, and the thermodynamic uncertainty relation between fluctuating current and entropy production \cite{Gingrich2016,Pietzonka2018,Li2019TUR}.
For  the adiabatic process in a Carnot cycle, the positive entropy production in the transition between two equilibrium states implies the process is in general irreversible unless the transition path is quasi-static, i.e., reversible and zero-entropy production process can only be achieved infinitely slow such that equilibrium is kept at every moment during the transition. Speeding up the transition from one equilibrium state to another in a time much shorter than the intrinsic relaxation time but still can reproduce the same output as a quasi-static process, has been a challenging issue, which requires non-trivial control protocols of one or more parameters \cite{Guery-Odelin2019a}.

For the past decade, this idea of designed accelerated transition has attracted attention in  quantum systems \cite{Ibanez2012,Campbell2017,Patra2017ShortcutsFields}, known as the shortcuts-to-adiabaticity. 
For  stochastic systems, it is possible to achieve a finite-rate transition between two designated equilibrium states \cite{Martinez2016,Chupeau2020,Plata2020} via a non-equilibrium path, and also to reduce the dissipated work \cite{cabara2020}.
Another recent advance is the shortcut-to-isothermality (ScI) transition \cite{Li2017,Li2019} in which instantaneous equilibrium (ieq) at a fixed temperature is maintained at all moments during the finite speed transition,  which was demonstrated experimentally in a Brownian particle under  a moving harmonic potential \cite{Albay2019} and trapping potentials of varying stiffness \cite{Albay2020,Albay2020a}. The  ScI transition manifests the idea of instantaneous equilibrium which generalizes the notion of equilibrium,  and a work relation for the symmetry of the ScI of forward and reverse processes was established \cite{Albay2021}.

The novel process of ScI protocol can achieve a finite-rate isothermal process while keeping the system at a fixed temperature during the transition for time-dependent protocols to vary the parameter(s) in a potential. However, in many theoretical and practical situations, one would like to heat up or cool down the system by varying the temperature in time. The role of a temperature change protocol is fundamentally different from the potential-parameter protocol and calls for the need to re-formulating the ieq process to examine its possibility and derive a new recipe. 
For instance, as demonstrated recently in the quasi-static adiabatic process in a trapped Brownian particle \cite{Martinez2015},  underdamped dynamics must be considered with time-dependent temperature protocol.
In this paper, we will theoretically formulate and derive the conditions for ieq under both time-dependent temperature and potential-parameter protocols for an underdamped Brownian particle under harmonic and non-harmonic trapping potentials. Simulations are then carried out to verify the theoretical results. This paper is organized as follows. In Sec. II, we define and develop the instantaneous equilibrium protocols for time-varying potential and temperature. 
Then the position, momentum, and stochastic energetics of the ieq process are calculated in Sec. III. 
Sec. IV introduces the ieq isentropic and zero entropy change processes, which are not  adiabatic  in contrast to the  corresponding processes under the quasi-static condition. 
In Sec. V, we generalize the work relation and derive a new  heat relation  for the ieq protocols of the forward and reverse processes.  We conclude and discuss the implications on micro heat engines and relevant experimental outlook in Sec. VI.

\section{Underdamped ieq  Dynamics under time-dependent potential and temperature protocols: the auxiliary escort potential} 
We consider an underdamped Brownian particle of mass $m$ and damping coefficient $\gamma$ moving in one dimension  under the potential $U_0(x,\lambda(t))$,
where $\lambda(t)$ is the time-dependent protocol that drives the potential. The time-dependent Hamiltonian of the system is given by 
\begin{equation}
H_0(x,p,\lambda(t))=\frac{p^2}{2m}+U_0(x,\lambda(t))
\end{equation}
where $p$ is the momentum of the particle. At the same time the temperature of the system is also changing whose protocol is given by the  inverse temperature $\beta(t)=1/T$ (the Boltzmann constant is set to unity for simplicity).
Similar to the ScI recipe \cite{Li2017} to achieve instantaneous equilibrium (ieq), the Ramp potential $U_0(x,\lambda(t))$ is escorted by a position and momentum-dependent
auxiliary potential $U_1(x,p,t)$ so that  the particle experiences a total Hamiltonian 
\begin{equation}
 H=H_0(x,p,\lambda(t))+ U_1(x,p,t)\label{U0U1udr}
\end{equation}
such that  the underdamped  Brownian particle will be in an ieq state obeying the Boltzmann distribution $\rho_{\rm ieq}(x,p,t)=e^{-\beta(t)(F(\lambda,\beta)- H_0(x,p,\lambda(t))}$,
where $ F(\lambda,\beta)=-\frac{1}{\beta(t)}\ln \int dp \int dx e^{-\beta(t) H_0(x,p,\lambda(t))}$ is the free energy of the Ramp(original) system at ieq for some instantaneous value of $\lambda$ and $\beta$ during the transition process.
The auxiliary potential $U_1(x,p,t)$ can be  determined for a given Ramp potential $U_0(x,\lambda(t))$ with the protocols $\lambda(t)$ and $\beta(t)$ from the following linear partial differential equation (see details of the derivations  in Appendix A)  
\begin{equation}
\dot{\beta}(F-H_0)+\beta\left(\dot{F} -\dot{\lambda}\frac{\partial U_0}{\partial \lambda}\right)=\gamma \frac{\partial^2 U_1}{\partial p^2}+\left(\frac{\partial U_0}{\partial x}-\frac{\gamma}{m} p\right)\beta  \frac{\partial U_1}{\partial p}-\beta \frac{p}{m} \frac{\partial U_1}{\partial x},\label{U1PDE}
\end{equation}
where $\dot{}$ denotes the derivative with respect to  time.
It is easy to see the $U_1(x,p,t)$ can be put into the form
\begin{equation}
 U_1(x,p,t)= \dot{\lambda}(t) f(x,p,\lambda(t),\beta(t))+\dot{\beta}(t) g(x,p,\lambda(t),\beta(t)).\label{U1fg}
\end{equation}
For a smooth switching on and off of the auxiliary potential, one usually imposes the conditions 
\begin{equation}
\dot{\lambda}(0)=\dot{\lambda}(\tau)=\dot{\beta}(0)=\dot{\beta}(\tau)=0\label{bc}, 
\end{equation}
which, however, are not required for the derivation of $U_1$. Here, $\tau$ is the duration of the transition process .
Since (\ref{U1PDE}) is linear in $U_1$ for general $U_0(x,\lambda(t))$, one can employ the power series expansion method to solve for $U_1$. 
Under the ieq protocol, the Brownian particle will experience the potential
$U_{\rm ieq}(x,p,t)\equiv U_0(x,\lambda(t))+U_1(x,p,t)$
and will be at ieq and obey the Boltzmann distribution $\rho_{\rm ieq}(x,p,t)$ at any time ($0\leq t \leq \tau$) during the transition process. In this paper, we focus on harmonic and non-harmonic trapping potentials of the form 
\begin{equation}
U_0=\dfrac{1}{2}\lambda(t)x^n,\quad n=2,4,6,\cdots\label{U0xn}
\end{equation} for explicit analytic calculations and numerical solutions of the corresponding underdamped Langevin dynamics. The time-dependent protocols
\begin{eqnarray}
\lambda(t)&=&\lambda(0)+\frac{\Delta \lambda}{2}\left(1-\cos\frac{\pi t}{\tau}\right),\quad \Delta \lambda\equiv \lambda(\tau)-\lambda(0)\label{lamproto}\\
T(t)&=&T(0)+\frac{\Delta T}{2}\left(1-\cos\frac{\pi t}{\tau}\right),\quad \Delta T\equiv T(\tau)-T(0) \label{Tpropto}
\end{eqnarray}
will be employed in the simulations of the Langevin dynamics.

  \subsection{The ieq auxiliary potential for $U_0=\frac{1}{2}\lambda(t) x^n$, $n=2,4,6\cdots.$ }
 For the Ramp potential given by (\ref{U0xn}), the Hamiltonian of the underdamped particle of mass $m$ is given by
\begin{equation}
H_0(x,p,\lambda(t))=\frac{p^2}{2m}+\frac{1}{2}\lambda(t) x^n, \quad n=2,4,6\cdots.\label{H0}
\end{equation}
  Here the expression for the auxiliary potential $U_1$ under time-dependent drivings of $\lambda(t)$ and $\beta(t)$ is given by (see detail derivations in Appendix A) 
\begin{equation}
 U_1(x,p,t)=  \frac{\tau_m\dot{\lambda}(t)}{n\lambda(t)} H_0(x,p-\gamma x,\lambda(t)) +  \frac{\tau_m\dot{\beta}(t)}{\beta(t)}\left[\frac{1}{2}H_0(x,p,\lambda(t))+\frac{1}{n}H_0(x,p-\gamma x,\lambda(t))\right],\label{U1udrxn}
\end{equation}
where $\tau_m\equiv m/\gamma$ is the inertia memory time of the underdamped particle. The underdamped particle experiences  the total potential
  \begin{equation}
U_{{\rm ieq}}=\frac{1}{2}\lambda(t) x^n+ U_1(x,p,t)
  \end{equation}
  and follows the ieq Boltzmann distribution  with the instantaneous free energy, given respectively by
  \begin{eqnarray}
    \rho_{\rm ieq}(x,p,t)&=& e^{\beta(t)[F(\lambda(t),\beta(t))-\frac{p^2}{2m}-\frac{\lambda(t)}{2} x^n] }\\
   \beta(t) F(\lambda(t),\beta(t))&=&\ln\left[\frac{n}{2\Gamma(\tfrac{1}{n})}\left( \frac{\beta(t)\lambda(t)}{2}\right)^{\frac{1}{n}}\sqrt{\frac{\beta(t)}{2\pi m}}\right]\label{Flambet}\\
&=&\tfrac{1}{n}\ln\lambda(t)+ (\tfrac{1}{2}+\tfrac{1}{n})\ln\beta(t)+\hbox{constant},
  \end{eqnarray}
where $\Gamma(x)$ is the Gamma function. Denote  the (time-dependent) ensemble average at ieq by $\langle \cdots \rangle\equiv \int dx  \int dp \cdots \rho_{\rm ieq}(x,p,t) $, direct calculations give
   \begin{equation}
  \langle p^2(t)\rangle = \frac{m}{\beta(t)};\quad \langle x^n(t)\rangle = \frac{2}{n\beta(t)\lambda(t)};\quad \langle x^2(t)\rangle=\left(\frac{2}{\beta(t)\lambda(t)}\right)^{2\over n} \frac{\Gamma({3\over n})}{\Gamma({1\over n})}.\label{meanx2}
  \end{equation} 
 
One can write the potential protocol in terms of the dimensionless function $\Lambda(t)$ as $\lambda(t)=\lambda(0) \Lambda(t)$, and express all energy scales in terms of the initial inverse temperature $\beta_0\equiv\beta(0)$. 
For $U_0$ of the form (\ref{U0xn}), one can express all lengths in terms of the natural spatial scale $\sigma\equiv (\beta_0\lambda(0))^{-\frac{1}{n}}$, and express time in unit of the  relaxation time $\tau_R\equiv \frac{\beta_0\gamma}{(\beta_0\lambda(0))^{\frac{2}{n}}}$. Then the Hamiltonian and potentials can be written using the dimensionless space, time, and momentum (${\tilde x}\equiv x/\sigma, {\tilde t}\equiv t/\tau_R, {\tilde p}\equiv p\tau_R/(m\sigma$)) variables. Hereafter, with the understanding that all space, time and momentum variables are expressed in their dimensionless form, the superscript $\tilde{}$ will be dropped for notation convenience. In terms of these dimensionless variables, the time-dependent Hamiltonian consists of kinetic ($K$) and potential energies, which are given by
\begin{eqnarray}
H&=&H_0+U_1, \quad H_0=K+  U_0, \quad \beta_0K= \tfrac{\alpha}{2} { p}^2; \quad \alpha\equiv \tfrac{\tau_m}{\tau_R} \\
 \beta_0U_0&=&\tfrac{1}{2}\Lambda({ t}) { x}^n\label{U0xn2}\\
  \beta_0U_1&=&\tfrac{\nu(t)}{2n}\left[(\alpha{ p}-{ x} )^2+\alpha \Lambda({ t}) { x}^n  \right]+  \tfrac{\dot{\beta}(t)}{4\beta(t)} \left[(\alpha{ p} )^2+\alpha \Lambda({ t}) { x}^n  \right];
  \quad \nu(t)\equiv \tfrac{\dot{\Lambda}(t)}{\Lambda(t)}+ \tfrac{\dot{\beta}(t)}{\beta(t)},\label{U1xpxn2} \end{eqnarray}
and (\ref{meanx2}) reads
\begin{equation}
    \alpha\langle p^2(t)\rangle = \frac{\beta_0}{\beta(t)};\quad \langle x^n(t)\rangle = \frac{2\beta_0}{n\beta(t)\Lambda(t)};\quad \langle x^2(t)\rangle=\left(\frac{2\beta_0}{\beta(t)\Lambda(t)}\right)^{2\over n} \frac{\Gamma({3\over n})}{\Gamma({1\over n})}.\label{meanx2b}
\end{equation} 
In these dimensionless units, the behavior of the underdamped Brownian particle depends only on the parameter $\alpha$ (the ratio of inertia memory and relaxation  times), the potential parameter $n$, the normalized protocol $\Lambda(t)$ and the relative inverse temperature protocol $\beta(t)/\beta_0$.
 
\subsection{Stochastic energetics and Thermodynamics of the ieq process}

The infinitesimal work ($dW$)  for the ieq trajectory from time $t$ to $t+dt$ can be calculated directly from (\ref{U0xn2}) and  (\ref{U1xpxn2}) to give
\begin{eqnarray}
    \beta_0dW&=&\beta_0\frac{\partial (U_0+U_1)}{\partial t} dt\\
    &=&\Big\lbrace \left[ \tfrac{1}{2}(1+ \tfrac{\alpha}{n}\nu(t) +\tfrac{\alpha}{2}\tfrac{\dot{\beta}(t)}{\beta(t)})\dot{\Lambda}(t)+ \tfrac{\alpha}{2}\Lambda(t)[ \tfrac{\dot{\nu}(t)}{n}+ \tfrac{1}{2} \tfrac{d}{dt} \left(\tfrac{\dot{\beta}(t)}{\beta(t)}\right) ] \right] x^n\label{dW}\\
  & &+   \tfrac{\dot{\nu}(t)}{2n} (\alpha p-x)^2 +  \tfrac{1}{4}\tfrac{d}{dt} \left(\tfrac{\dot{\beta}(t)}{\beta(t)}\right) (\alpha p)^2 \Big\rbrace  dt.\nonumber
  \end{eqnarray}  
The energy change along the ieq trajectory from $t$ to $t+dt$ is given by energy conservation  as  
\begin{equation}
 \beta_0dE_{\rm ieq}= \tfrac{\alpha}{2}dp^2 +\beta_0(dU_0+dU_1).
\end{equation}
The heat can then be computed using the First Law of Thermodynamics, $dQ=dE_{\rm ieq}-dW$, but may be more intuitive to derive it from the definition of stochastic heat (see Appendix B for details) to give
\begin{equation}
 \beta_0dQ= \tfrac{\alpha}{2}dp^2 +\beta_0\tfrac{\partial (U_0+U_1)}{\partial x}\circ dx + \beta_0\tfrac{\partial U_1}{\partial p}\circ dp\label{dQ},
\end{equation}
where $\circ$ denotes the Stratonovich calculus convention. (\ref{dQ}) is more convenient for measuring the heat in simulations or experiments which record the stochastic trajectories. 
  
One can  define the instantaneous entropy of the system, $S(t)\equiv\frac{\partial F}{\partial T}=\beta^2(t) \frac{\partial F}{\partial \beta}$.
Thus  during the ieq  transition, the instantaneous free energy, entropy, total energy, and potential energy ($F(t)$, $S(t)$, $\langle E_0(t)\rangle$, $\langle U_0(t)\rangle$) are well-defined state functions (functions that depend only on the state parameters $\lambda$ and $\beta$ at time $t$). It is easy to show that 
\begin{equation}
  \langle E_0(t)\rangle\equiv \langle H_0(t)\rangle=F(t)+T(t)S(t), \label{E0FTS}
\end{equation} manifesting the ieq properties.
Furthermore, $\langle E_{\rm ieq}(t)\rangle=\langle E_0(t)\rangle+\langle U_1(t)\rangle$, revealing that additional energy input, $\langle U_1(t)\rangle$, is required to maintain the ieq.
   
It is worth noting that for smooth protocols satisfying  (\ref{bc}), $\Delta  U_1\equiv \langle U_1(\tau)\rangle - \langle U_1(0)\rangle=0$, and hence $\Delta E_{\rm ieq}=\Delta E_0= \Delta F +\Delta (TS)$. Invoking the First Law of Thermodynamics, $\Delta E_{\rm ieq}=\langle Q 
\rangle +\langle W \rangle$, and  using $\Delta (TS)= \int_0^{\tau}T(t)\dot{S}(t)dt + \int_0^{\tau}S(t)\dot{T}(t)dt$, one gets
\begin{equation}
    \langle W \rangle-\left(\Delta F+\int_0^{\tau}S(t)\dot{T}(t)dt \right)=\int_0^{\tau}T(t)\dot{S}(t)dt- \langle Q 
    \rangle.\label{WdissQdiss}
\end{equation}
One can easily show that the quantity
\begin{equation}
 \Delta \Phi\equiv \Delta F+\int_0^{\tau}S(t)\dot{T}(t)dt=\int_0^{\tau}dt\dot
 {\lambda}\frac{\partial F}{\partial \lambda}=\int_0^{\tau}dt\left\langle\frac{\partial H_0}{\partial t}\right\rangle\label{Phidef}
\end{equation}
which can be interpreted as the  mean work in the corresponding Ramp process in the quasi-static limit (reversible). Hence the difference $\langle W \rangle-\Delta \Phi\equiv W_{\rm diss}$ is the dissipated (or irreversible) work of the ieq process.
In a similar spirit, $T\dot{S}dt$ is the instantaneous heat (the reversible part) production  for the Ramp process in the quasi-static limit. 
Thus, $\int_0^{\tau}T(t)\dot{S}(t)dt- \langle Q \rangle\equiv -Q_{\rm diss}$ in the RHS (\ref{WdissQdiss}) can be interpreted as the dissipated heat that flows out of the system for the ieq transition.  (\ref{WdissQdiss}) implies that dissipated work on the system equals the dissipated heat that flows out of the system for the ieq process, i.e., $ W_{\rm diss}=-Q_{\rm diss}$, which can be viewed as a generalization of the equilibrium process in   which both $W_{\rm diss}$ and $Q_{\rm diss}$ vanish.
 
\section{Position, momentum, and energetic statistics of ieq process under time-dependent temperature and stiffness changes}

Under ieq, the system will obey Boltzmann distribution of the Hamiltonian (\ref{H0}) with the distribution $\rho_{\rm ieq}(x,p,t)$ factorized into the position distribution and a Gaussian momentum distribution: \begin{eqnarray}
\rho_{\rm ieq}(x,p,t)&=&P(x,t)P_G(p,\beta(t))\\
P_G(p,\beta(t))&\equiv& \sqrt{\frac{\alpha\beta(t)}{2\pi\beta_0}}e^{-\frac{\alpha\beta(t)}{2\beta_0}p^2}.\label{PGpt}
\end{eqnarray}
Remarkably, the momentum distribution takes a rather universal Gaussian form that is independent of $U_0$ (and hence independent of the potential protocol $\lambda(t)$). Furthermore, since the mean kinetic energy $\beta_0\langle K \rangle = \alpha \langle p^2\rangle/2$ and  $\langle p^2\rangle$  is given by the variance of $P_G(p,\beta)$ in (\ref{PGpt}), one has the general universal result for the mean instantaneous kinetic energy under ieq:
\begin{equation}
\langle K(t) \rangle = \tfrac{1}{2} T(t)\label{meanKt},
\end{equation}
which depends only on the temperature protocol  and is independent of $\alpha$, i.e., even in the over-damped limit of $\alpha\to 0$. There are always kinetic energy changes associated with temperature variations.
(\ref{meanKt}) can be viewed as an instantaneous thermal energy partition principle for the Brownian particle  under instantaneous temperature $T(t)$, manifesting the nature of the ieq process.

\begin{figure}
\subfigure[]{ \includegraphics[width=\figwss]{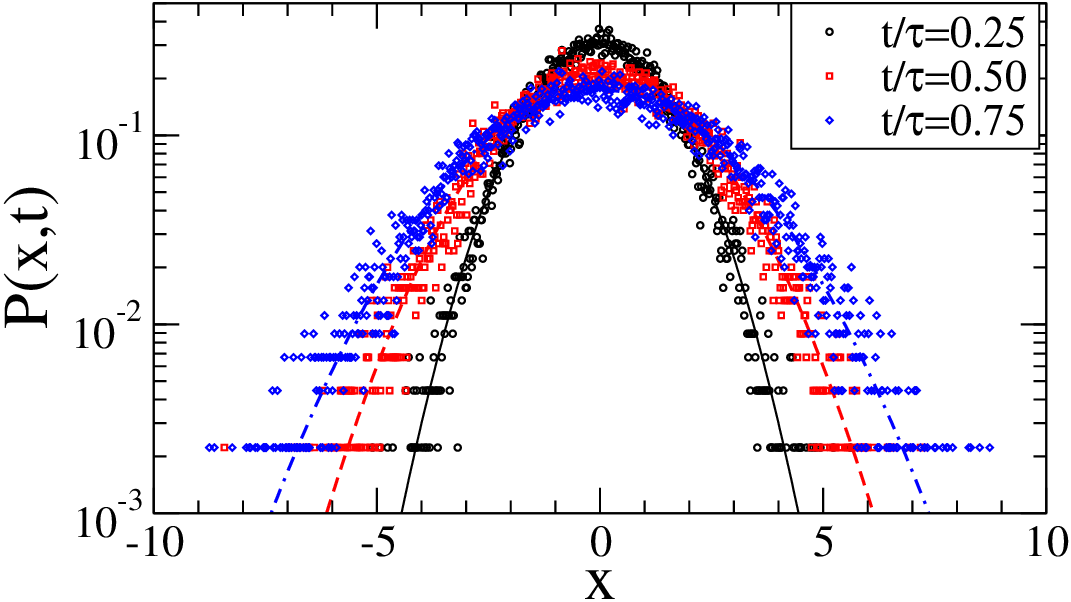}}
\subfigure[]{ \includegraphics[width=\figwss]{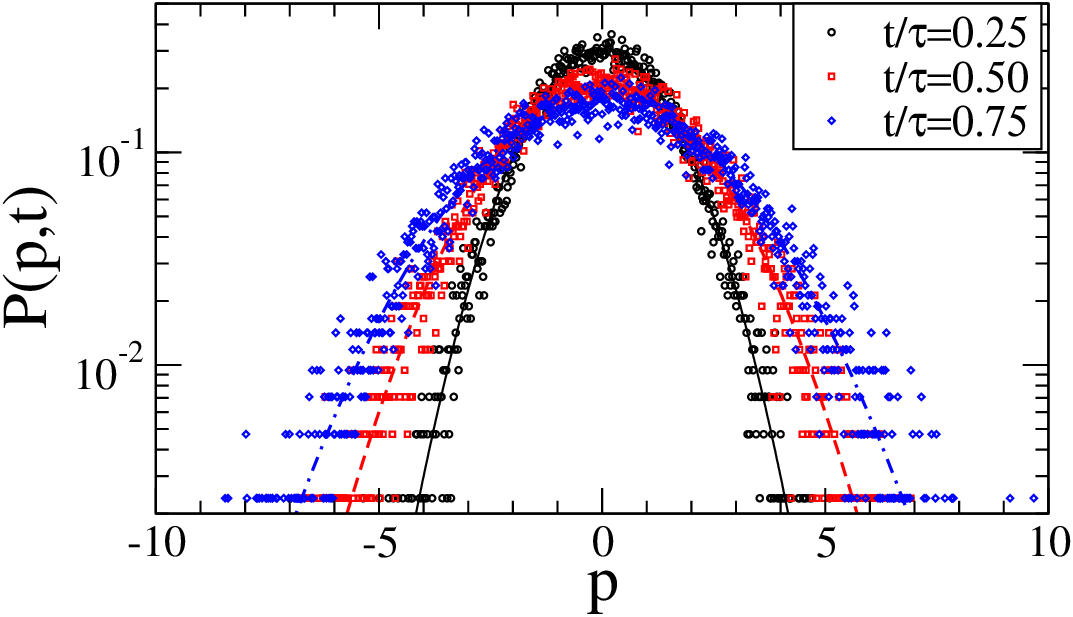}}
\subfigure[]{ \includegraphics[width=\figwss]{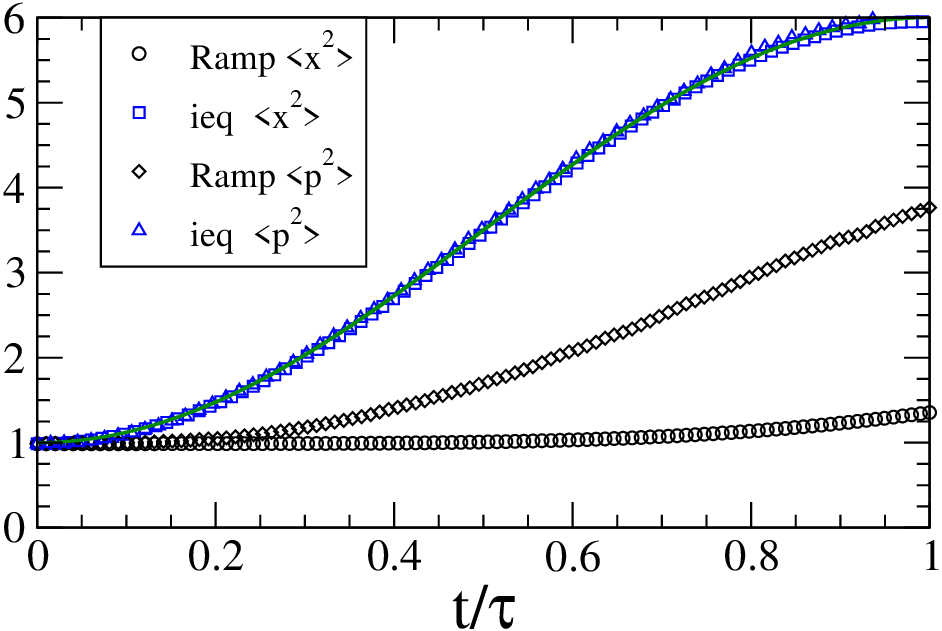}}
\subfigure[]{ \includegraphics[width=\figwss]{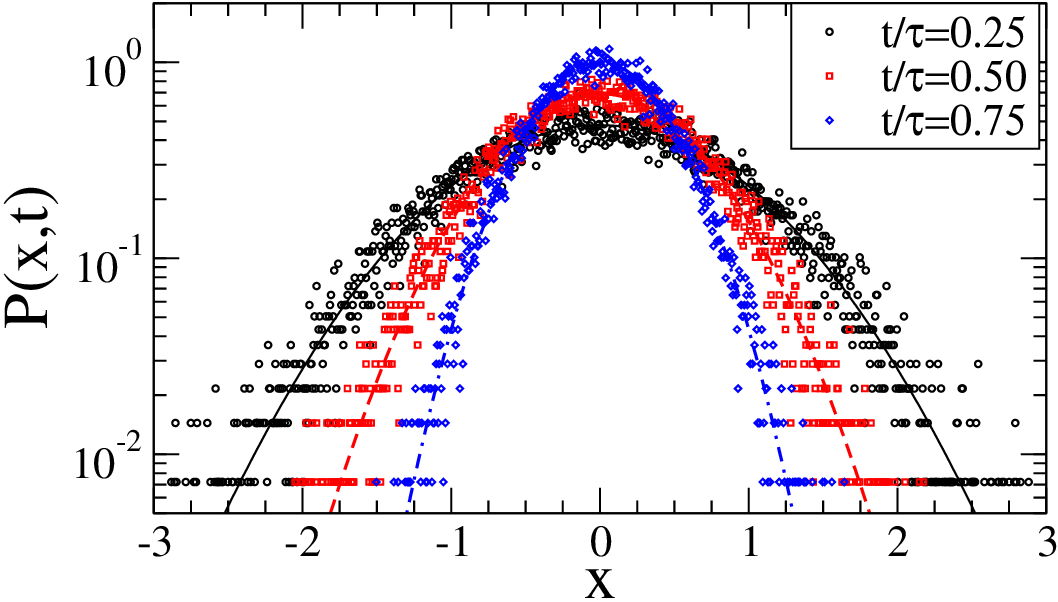}}
\subfigure[]{ \includegraphics[width=\figwss]{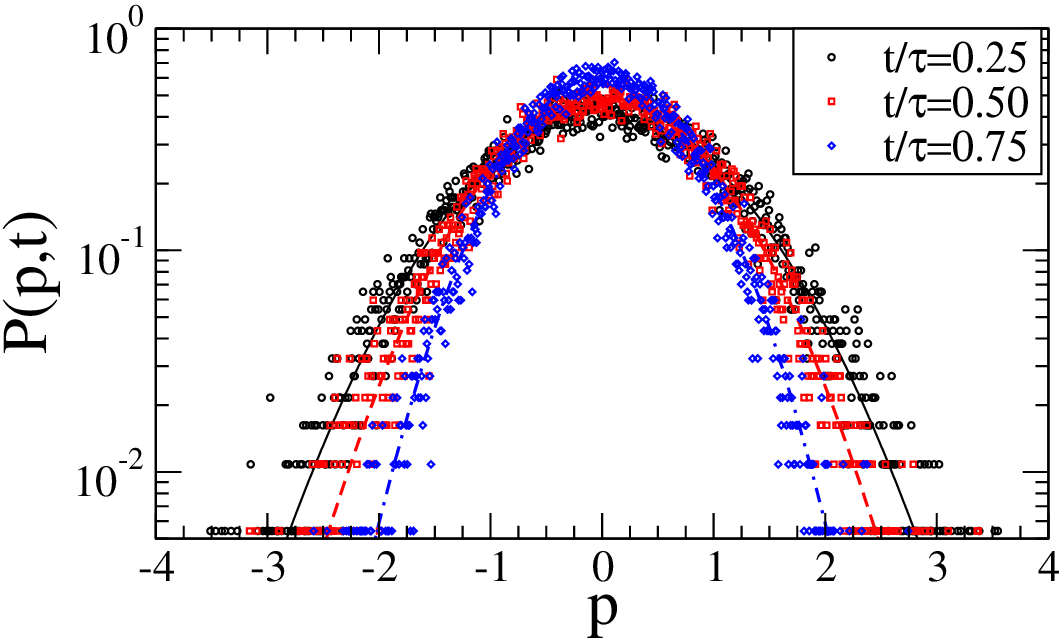}}
\subfigure[]{ \includegraphics[width=\figwss]{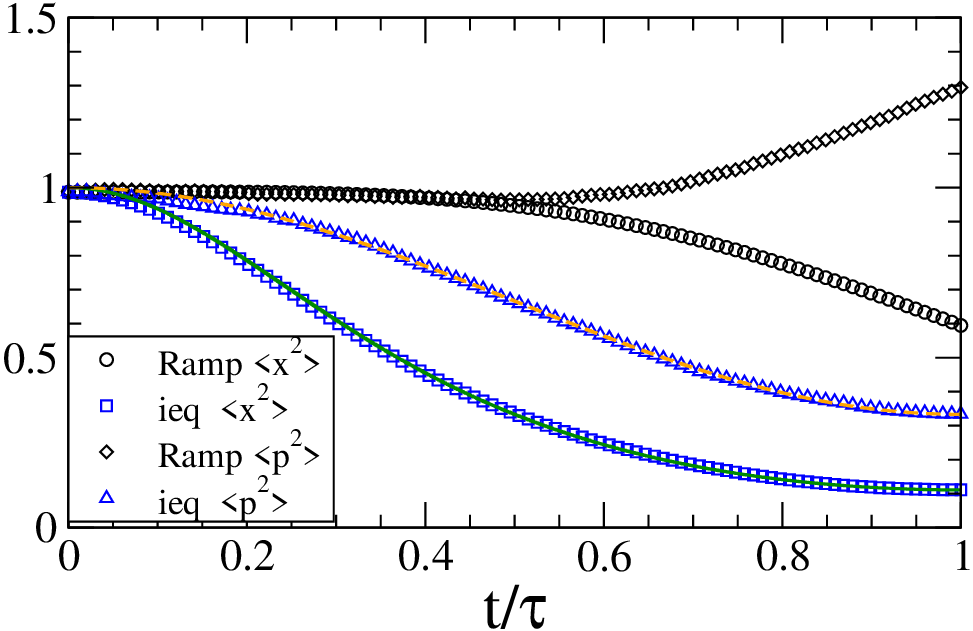}}
\subfigure[]{ \includegraphics[width=\figwss]{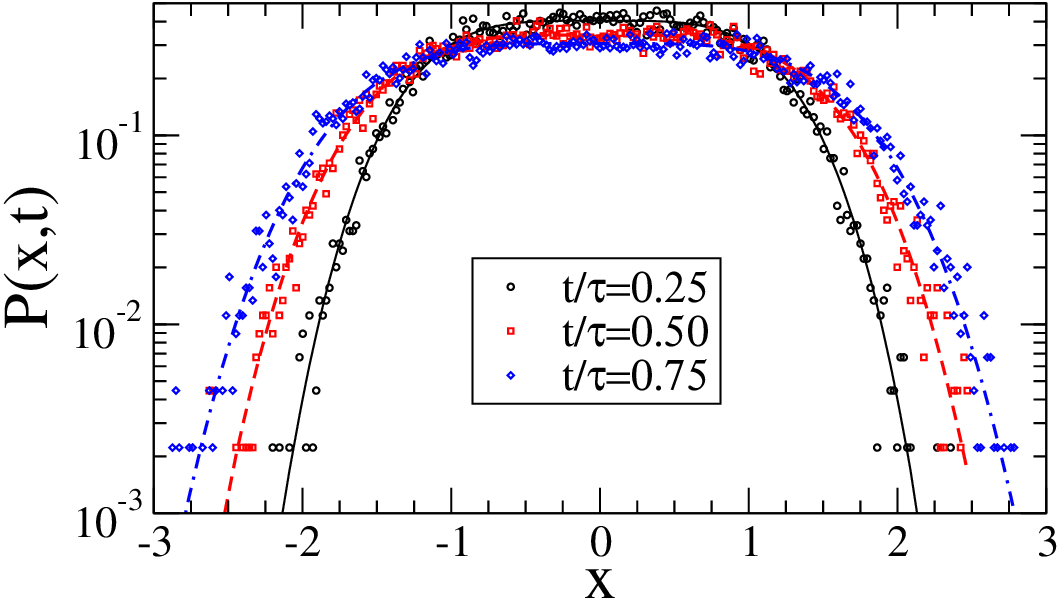}}
\subfigure[]{ \includegraphics[width=\figwss]{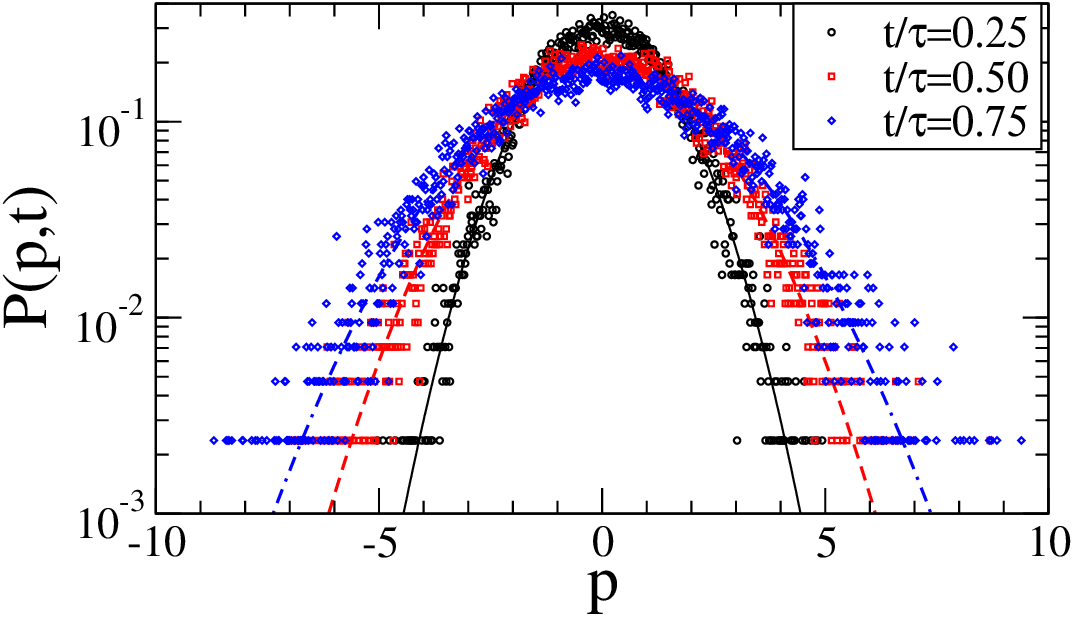}}
\subfigure[]{ \includegraphics[width=\figwss]{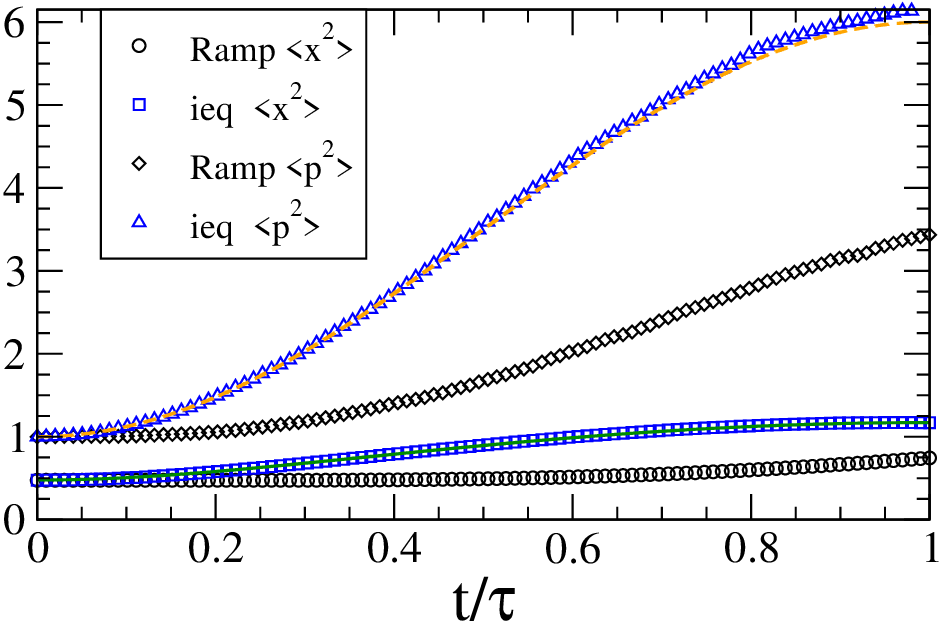}}
\caption{Position and momentum statistics of an underdamped Brownian particle under ieq protocol for the  potential $U_0(x,\lambda(t))=\frac{1}{2}\lambda(t)x^n$ for $n=2$and $n=4$, under protocols (\ref{lamproto}) and (\ref{Tpropto}). (a) Time-dependent position distributions $P(x,t)$ for three different times during the transition for the case of pure heating with $T(\tau)/T(0)=6$ while the stiffness is kept constant at unity. The solid curves are the corresponding Boltzmann distributions verifying the ieq nature. (b) Time-dependent momentum distributions $P(p,t)$ for three different times  in (a). (c) The position and momentum variances for the case in (a)  plotted against the transition period $\tau$ for the ieq process. The variances for the Ramp  process are also shown for comparison. The curves show the respective theoretical results. (d) Similar to (a) for $T(\tau)/T(0)={1\over 3}$ and the stiffness increases as $\lambda(\tau)/\lambda(0)=3$. (e) Similar to (b) for the case in (d). (f) Similar to (c) for the protocols in (d). (g) Position distributions with the same protocols as in (a) but for $n=4$ non-harmonic potential. (h) Momentum distributions  for the case in (g). (i) The position and momentum variances for the case in (g)  plotted against the transition period $\tau$ for the ieq process for  the non-harmonic potential.}
\label{xsqtudr}
\end{figure}

Here we derive analytic results for the ieq process under the Ramp potential of the form (\ref{U0xn2}).
Direct calculation gives the ieq position distribution
\begin{equation}
P(x,t)=\frac{n}{2\Gamma({1\over n})}\left(\frac{\beta(t)\Lambda(t)}{2\beta_0} \right)^\frac{1}{n}e^{-\frac{\beta(t)\Lambda(t)}{2\beta_0}x^n}.\label{Pxt}
\end{equation}
The position and momentum distributions of an underdamped Brownian particle under harmonic and non-harmonic Ramp potentials  with  various protocols at different times measured from Langevin dynamics simulations are shown in Fig. \ref{xsqtudr}. 
The momentum distributions  (Fig. \ref{xsqtudr}b,e,h) are indeed zero-mean Gaussian with a time-dependent variance that follows the instantaneous temperature $T(t)$ but independent of $\lambda(t)$ and $U_0$, as given by (\ref{PGpt}) (curves). 
On the other hand, the position distributions (Fig. \ref{xsqtudr} a,d,g) faithfully follow the Boltzmann form of $e^{-\beta(t) U_0(x,\lambda(t))}$ at all times for the harmonic and non-harmonic $U_0$ as given by (\ref{Pxt}). The time evolution of the variances of position and momentum during the entire ieq transition are also shown (Fig. \ref{xsqtudr}c,f,i), indicating that both variances are determined by the instantaneous values of $\beta$ and $\lambda$ that agree perfectly with the theoretical results in (\ref{meanx2b}). 
The instantaneous position and momentum variances under the corresponding Ramp protocols are also displayed for comparison. Under the Ramp protocols, the position and momentum of the Brownian particle respond much slower than the relatively fast transition rate due to both the inertia ($\tau_m$) and viscous ($\tau_R$) effects, and will relax to the final equilibrium state at a time much later than $\tau$.

\begin{figure}
\subfigure[]{ \includegraphics[width=3.in]{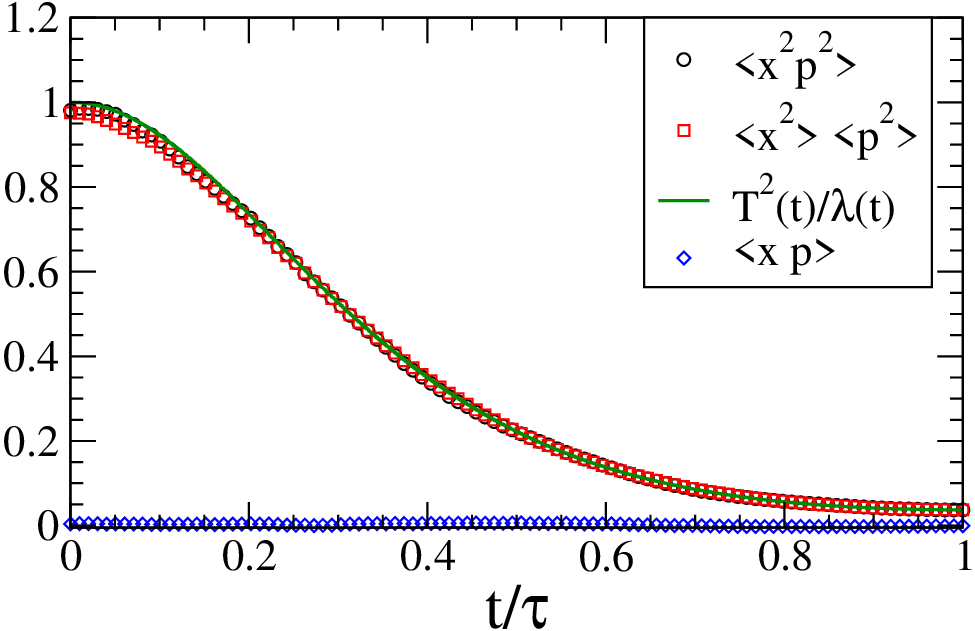}}
\subfigure[]{ \includegraphics[width=3.in]{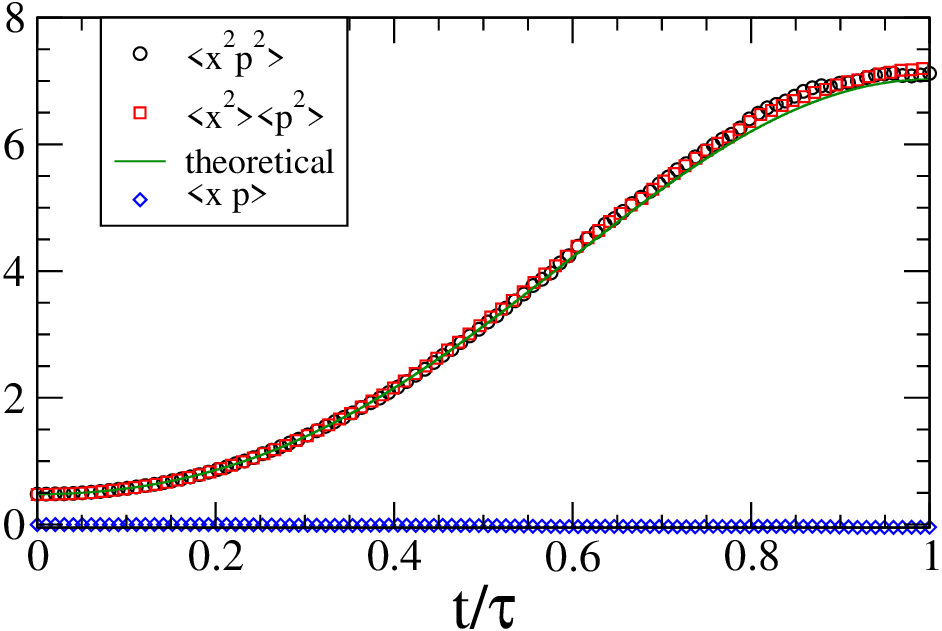}}
\caption{  Equal-time cross-correlations of the position and momentum plotted as a function of time during the ieq process to illustrate that the position and momentum distributions are independent. (a) Harmonic potential with  the protocols as in Fig. \ref{xsqtudr}d. The solid curve, $T^2(t)/\lambda(t)$ is the theoretical result of  $\langle x^2(t)\rangle  \langle p^2(t)\rangle$. (b) $n=4$ non-harmonic potential with the protocols as in Fig. \ref{xsqtudr}g. The theoretical result of  $\langle x^2(t)\rangle  \langle p^2(t)\rangle$ given by  (\ref{meanx2b}) is shown by the solid curve. }
\label{xvcorr}
\end{figure}

To further verify that the position and momentum are indeed distributed independently, the cross-correlations of $x$ and $p$ are measured in the simulations for harmonic (Fig. \ref{xvcorr}a) and non-harmonic ($n=4$, Fig. \ref{xvcorr}b) potentials under different protocols. The results indeed show clearly that  $ \langle x^2(t) p^2(t)\rangle=\langle x^2(t)\rangle  \langle p^2(t)\rangle$ and  $\langle x(t) p(t)\rangle=\langle x(t)\rangle  \langle p(t)\rangle=0$, confirming the $x$ and $p$ are independently distributed under ieq processes. In addition, $\langle x^2(t)\rangle  \langle p^2(t)\rangle$ shows perfect agreement with the analytic results (solid curve) in (\ref{meanx2b}).
 
\begin{figure}
\subfigure[]{ \includegraphics[width=3.in]{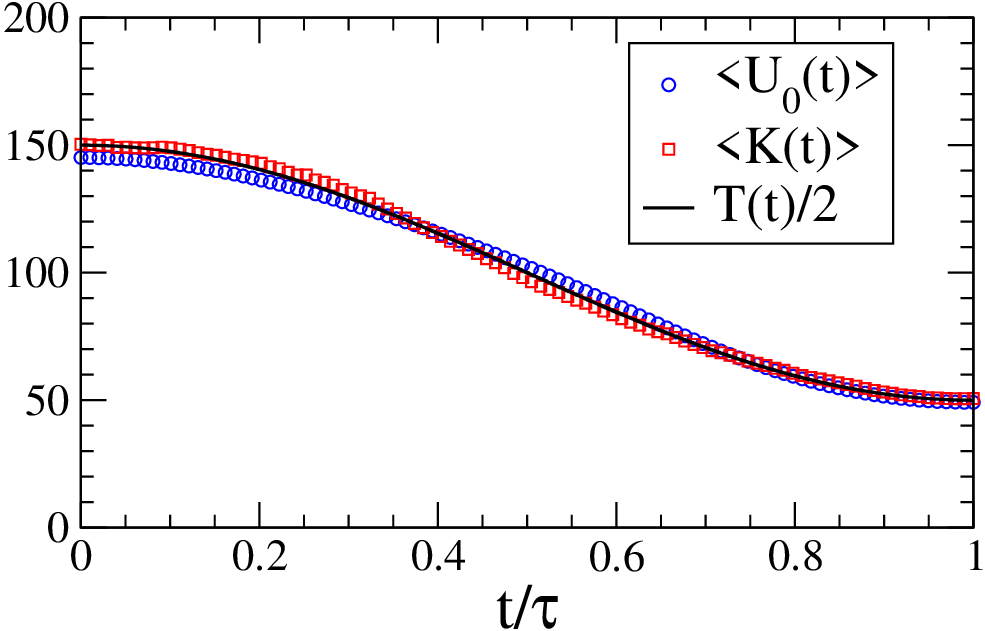}}
\subfigure[]{ \includegraphics[width=3.in]{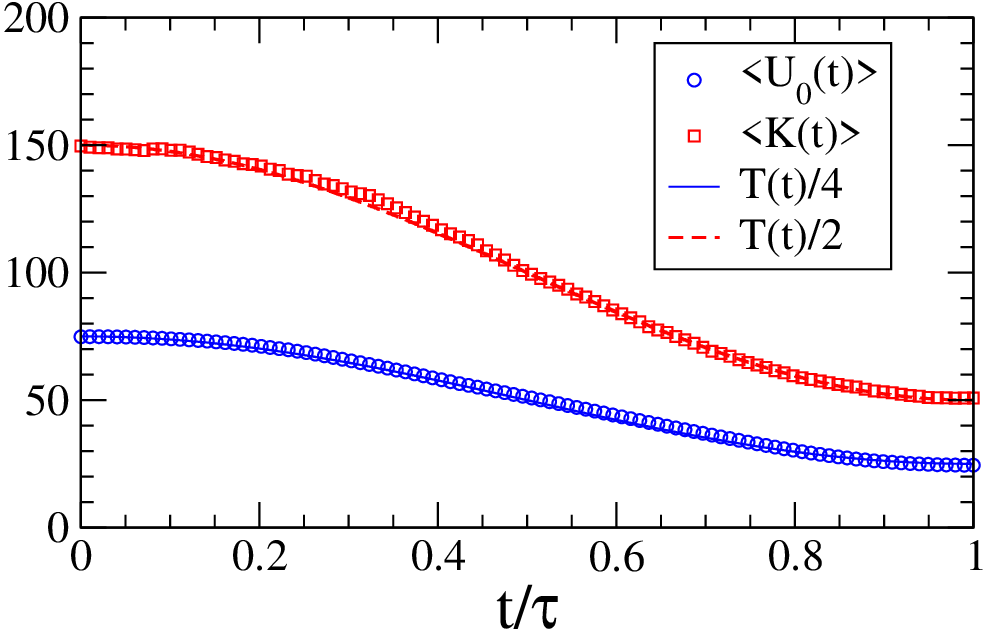}}
\caption{Time variation of the mean potential and kinetic energies  of an underdamped Brownian particle under ieq protocol for the potential $U_0(x,\lambda(t))=\frac{1}{2}\lambda(t)x^n$ under protocols (\ref{lamproto}) and (\ref{Tpropto}) with $T(\tau)/T(0)={1\over 3}$ and the $\lambda(\tau)/\lambda(0)=3$. (a) For the harmonic potential $n=2$. The solid curve shows the theoretical result of $T(t)/2$ for both $\langle U_0(t)\rangle$ and  $\langle K(t)\rangle$. (b)  For the non-harmonic potential $n=4$. The solid and dashed curves show the theoretical results of $T(t)/4$ and $T(t)/2$ for  $\langle U_0(t)\rangle$ and  $\langle K(t)\rangle$ respectively.}
\label{U0Kudr}
\end{figure}

With (\ref{meanx2b}), the instantaneous mean energies and entropy along the ieq path can be easily calculated to give
\begin{eqnarray}
    \langle K(t) \rangle& =&  \tfrac{1}{2\beta(t)},\quad \langle U_0(t) \rangle =  \tfrac{1}{n\beta(t)},\quad  \langle E_0(t) \langle\equiv \rangle H_0(t) \rangle = (\tfrac{1}{2} +\tfrac{1}{n}) \tfrac{1}{\beta(t)}\label{equiE}\\
    \langle E_{\rm{ieq}}(t) \rangle&\equiv&\langle H(t) \rangle = (\tfrac{1}{2} +\tfrac{1}{n}) \tfrac{1}{\beta(t)}+\langle U_1(t) \rangle \\
    \langle U_1(t) \rangle &=& \frac{\alpha(\alpha+\frac{2}{n})}{2\beta(t)}\left(\frac{\nu(t)}{n}+ \frac{\dot{\beta}(t)}{2\beta(t)}\right)+ \frac{\Gamma({3\over n})}{\Gamma({1\over n})}\frac{\nu(t)}{2n\beta_0}\left(\frac{2\beta_0}{\beta(t)\Lambda(t)}\right)^{2\over n}\\
    S(t)&=& -\tfrac{1}{n}\ln \Lambda(t) - (\tfrac{1}{2} +\tfrac{1}{n})\ln \tfrac{\beta(t)}{\beta_0} +\hbox{constant}\label{St}\\
    F(t)&=& \langle E_0(t) \rangle-T(t)S(t).
\end{eqnarray}
It is worth noting that (\ref{equiE}) can be interpreted as an instantaneous `` equi-partition" energy principle (the  energy $\langle E_0\rangle$ is equally partitioned into kinetic and potential energies only for the harmonic potential case) for the particle under $H_0$, echoing the nature of the ieq process.
Fig. \ref{U0Kudr} shows the instantaneous mean kinetic and potential energies in the ieq processes measured in Langevin simulations (symbols) for harmonic (Fig. \ref{U0Kudr}a) and non-harmonic ($n=4$, Fig. \ref{U0Kudr}b) potentials under  different protocols. The corresponding theoretical results from (\ref{equiE}) are also displayed (curves), showing very good agreement.

We further consider the change of energies for the whole ieq process from $t=0$ to $t=\tau$. For smooth switching of the time-dependent protocols, the boundary conditions (\ref{bc})  hold and the auxiliary potential vanishes at $t=0$ and $t=\tau$.
The change in the kinetic and auxiliary potential  energies are universally (i.e., independent of $U_0$) given by
  \begin{equation}
\langle \Delta K \rangle =\tfrac{1}{2}\Delta T, \quad \langle \Delta U_1\rangle=0; \qquad \Delta T\equiv T(\tau)-T(0).\label{meanE}
\end{equation}
And for $U_0$ of the form (\ref{U0xn}),
the energy change is easily computed to give
\begin{equation}
\langle \Delta E_{\rm ieq} \rangle = \langle \Delta E_0 \rangle =(\tfrac{1}{2}+\tfrac{1}{n})\Delta T, \quad \langle \Delta U_0 \rangle =\tfrac{1}{n}\Delta T.\label{meanE2}
  \end{equation}
Notice that the above energies depend only on the final and initial temperature difference and are independent of $\alpha$, $\tau$, and the protocols $\lambda(t)$ and $\beta(t)$.
 
\begin{figure}
    \subfigure[]{ \includegraphics[width=3.in]{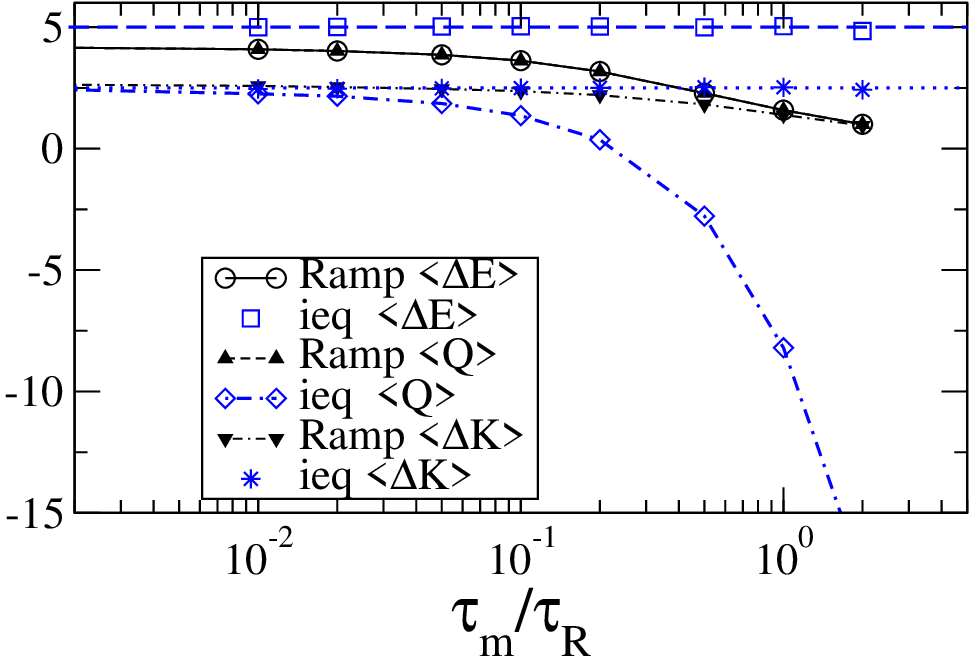}}
    \subfigure[]{ \includegraphics[width=3.in]{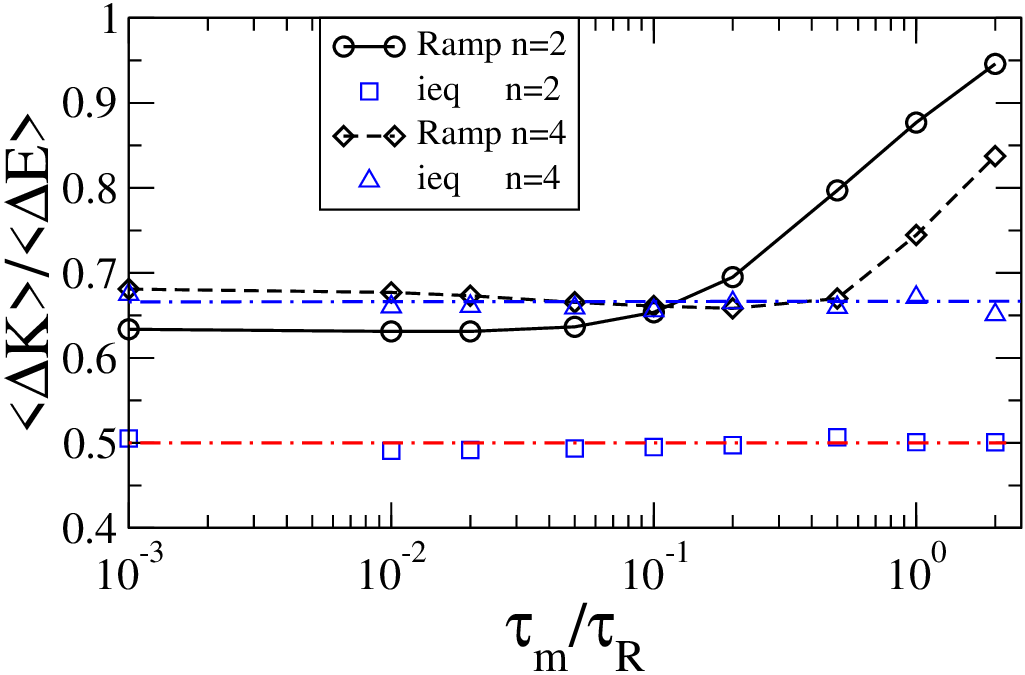}}
    \caption{(a) Mean energy and heat change plotted as a function of $\alpha\equiv\tau_m/\tau_R$ of an underdamped Brownian particle under ieq and Ramp  protocols with $\tau=\tau_R$, for the harmonic potential $U_0(x,\lambda(t))=\frac{1}{2}\lambda(t)x^2$ under protocols (\ref{lamproto}) and (\ref{Tpropto}) with $T(\tau)/T(0)=6$ and  $\lambda$ kept constant.  (b) Mean fraction of kinetic energy  plotted as a function of $\tau_m/\tau_R$ of an underdamped Brownian particle under ieq and Ramp  protocols for the harmonic and non-harmonic ($n=4$) potentials. The protocols are the same as in (a). }
    \label{EQKudr}
\end{figure}

The effect of inertia is revealed as the parameter $\alpha$ is varied. Fig.\ref{EQKudr}a shows the energy change and heat into the system for the protocol of pure heating under ieq and Ramp transitions. Since $\lambda$ does not change in this case and there is no work for the Ramp protocol, hence $\langle \Delta E \rangle=Q$ for the Ramp case. 
For the corresponding ieq process, $\langle \Delta E \rangle$ is independent of $\alpha$ with a value in good agreement with (\ref{meanE}) (dashed horizontal line) for given initial and final temperatures. However, extra work is needed to maintain ieq, resulting in heat dissipation whose magnitude increases with $\alpha$ (more work is required to maintain a more massive Brownian particle to stay at ieq).  
The mean kinetic energy change is also independent of $\alpha$  with the same values for both harmonic and non-harmonic potentials, which agrees perfectly with (\ref{meanE}). 
In particular, even  for the over-damped limit of $\alpha\to 0$, there is a significant  kinetic energy contribution  for both the Ramp and ieq processes, exemplifying the fact that the momentum degree of freedom (underdamped dynamics) is essential for any temperature changing protocol\cite{Martinez2015}.
 
Furthermore, the fractions of kinetic and potential energy contributions in the ieq process,
\begin{equation}
 \frac{\langle \Delta K \rangle}{\langle \Delta E_{\rm ieq} \rangle}=\frac{n}{n+2},\quad \frac{\langle \Delta U_{\rm ieq} \rangle}{\langle \Delta E_{\rm ieq} \rangle}=\frac{2}{n+2}\label{KEratio}
\end{equation}
are constants and independent of the initial and final temperatures.  Fig.\ref{EQKudr}b shows the fraction of kinetic energy contribution to the total energy change  with  the time-dependent protocols of both $\lambda$ and $\beta$.
For ieq processes, the ratio is a constant and independent of $\alpha$, and agrees well  with (\ref{KEratio}) (horizontal dot-dashed lines). 
On the other hand, for the Ramp process, the kinetic and potential energy changes depend on the protocols, the duration $\tau$, and $\alpha$. ${\langle \Delta K \rangle}/{\langle \Delta E \rangle}$ remains a finite fraction for $\alpha\to 0$ indicating the necessity of underdamped dynamics even for the over-damped case in both ieq and Ramp processes with temperature changes.

The entropy change of the system for the ieq process can be easily calculated to give
\begin{equation}
 \Delta S= -\tfrac{1}{n}\ln \Lambda(\tau) +(\tfrac{1}{2} +\tfrac{1}{n})\ln \tfrac{T(\tau)}{T(0)}\label{DeltaS}.
\end{equation}
For pure heating under ieq, $\Delta S>0$ as expected, but the entropy change can be negative upon simultaneous strong compression. More details on the properties of entropy under the ieq process will be investigated in the next section.
 
\begin{figure}
\subfigure[]{ \includegraphics[width=3.in]{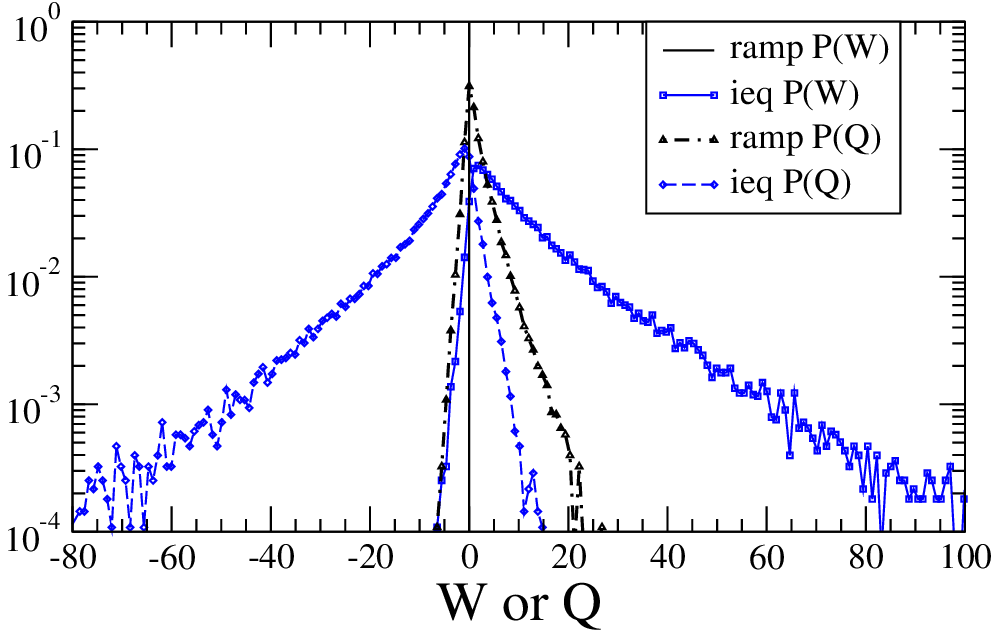}}
\subfigure[]{ \includegraphics[width=3.in]{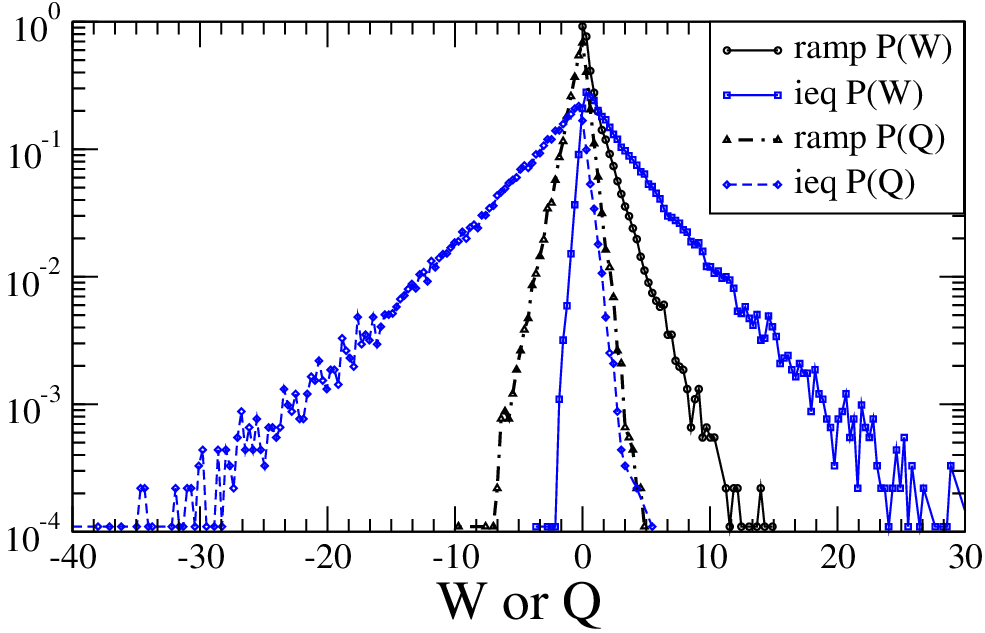}}
\caption{Work and heat distributions of an underdamped Brownian particle under ieq and Ramp  protocols for the harmonic potential $U_0(x,\lambda(t))=\frac{1}{2}\lambda(t)x^2$ under protocols (\ref{lamproto}) and (\ref{Tpropto})$W$ and $Q$ are in units of $T(0)$. (a) The protocol of pure heating  with $T(\tau)/T(0)=6$ and $\lambda$ kept constant.  There is no work for the Ramp case and the distribution is a $\delta-$function represented by the vertical straight line at $W=0$. (b) For protocols with $T(\tau)/T(0)={1\over 3}$ and  $\lambda(\tau)/\lambda(0)=3$.  }
\label{PWPQudr}
\end{figure}

\begin{figure}
\subfigure[]{ \includegraphics[width=3.in]{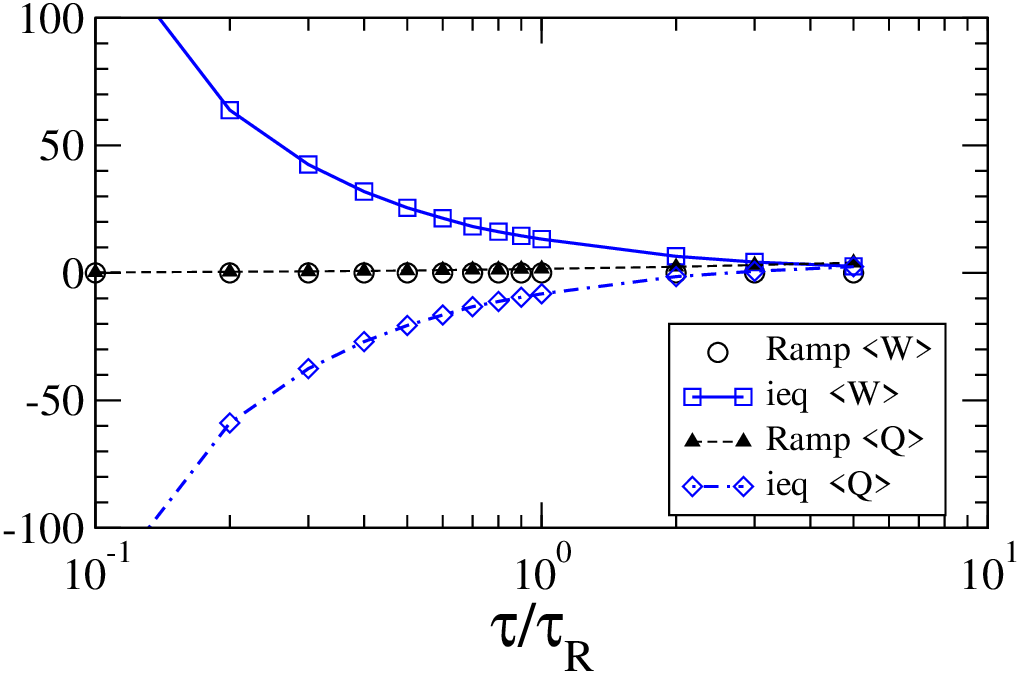}}
\subfigure[]{ \includegraphics[width=3.in]{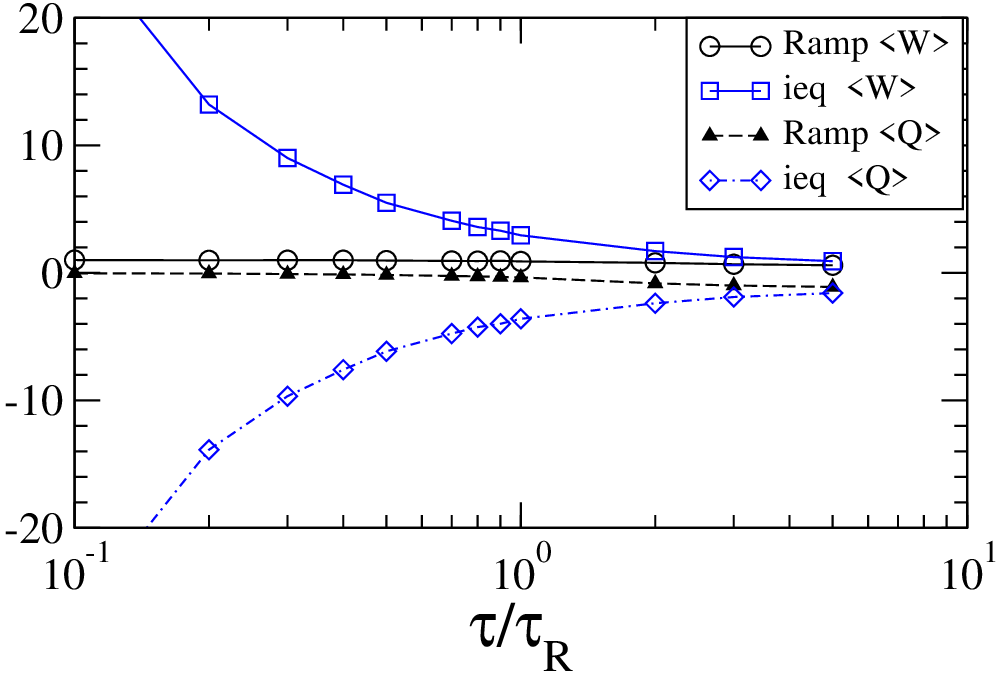}}
\caption{Mean work and heat plotted as a function of the transition period $\tau$ of an underdamped Brownian particle under ieq and Ramp  protocols for the harmonic potential $U_0(x,\lambda(t))=\frac{1}{2}\lambda(t)x^2$. (a) The case of pure heating with $T(\tau)/T(0)=6$ with $\lambda$ kept constant. There is no work for the Ramp case. The curves show the respective theoretical results. (b) Similar to (a) but for the case of  $T(\tau)/T(0)={1\over 3}$ and  $\lambda(\tau)/\lambda(0)=3$. }
\label{meanQW}
\end{figure}

The stochastic work and heat in an ieq trajectory can be measured  using (\ref{dW}) and (\ref{dQ}), and their distributions are shown in
Fig. \ref{PWPQudr} for  harmonic $U_0$ under two sets of protocols. The work and heat distributions for the  corresponding Ramp processes are also measured  for comparison. 
In general, the distribution of ieq work is much broader than that of the Ramp case due to the extra work done on the system by the auxiliary potential $U_1$ to maintain ieq. For example, there is no work in the Ramp transition for the case of pure heating up (Fig. \ref{PWPQudr}a), but the distribution is broad and has a long tail for the corresponding ieq transition. The ieq heat distribution is also very broad and skewed strongly to the negative values, indicating  large heat flow (dissipated) out of the system.
The  mean work in the ieq process can be computed theoretically by taking the ensemble average of (\ref{dW}) and using (\ref{meanx2}) to give
\begin{eqnarray}
\beta_0\langle W\rangle&=&\int_0^\tau dt \Bigg\langle  \frac{\partial\beta_0 (U_0+U_1)}{\partial t}\Bigg\rangle\nonumber\\
& =&\int_0^\tau dt  \tfrac{{\beta_0}}{\beta(t)}\left[ \tfrac{1}{n}(1+\tfrac{\alpha\nu(t)}{n}+\tfrac{\alpha}{2} \tfrac{\dot{\beta}(t)}{\beta(t)}) \tfrac{\dot{\Lambda}(t)}{\Lambda(t)} +\alpha(\tfrac{1}{2}+\tfrac{1}{n}) \left( \tfrac{\dot{\nu}(t)}{n}+(\tfrac{\dot{\beta}(t)}{\beta(t)})^2 \right)\right]\nonumber\\
& &+  \tfrac{\Gamma({3\over n})}{n^2\Gamma({1\over n})}\int_0^\tau  \nu^2(t) \left( \tfrac{2\beta_0}{\beta(t)\Lambda(t)}\right)^{\frac{2}{n}} dt,\label{meanW}
\end{eqnarray}
in which the integrals can be evaluated numerically. From (\ref{meanE2}) and (\ref{meanW}), the theoretical mean heat can be obtained from the First Law of Thermodynamics 
\begin{equation}
\langle Q\rangle=(\tfrac{1}{2}+\tfrac{1}{n})\Delta T- \langle W\rangle.\label{meanQ}
\end{equation}  Fig. \ref{meanQW} shows the mean heat and work for the ieq transition as a function of the transition duration $\tau$. The results for the corresponding Ramp transition are also displayed for comparison.  For fast transitions (small $\tau$), much more work is done on the system under ieq, and at the same time a large amount of heat is dissipated (flow out). As $\tau$ becomes large, the mean heat and mean work of both the Ramp and ieq processes approach  their quasi-static values. The mean work and heat under ieq agree well with the theoretical values given by (\ref{meanW}) (solid curve)  and (\ref{meanQ}) (dashed curve), respectively.

\section{Instantaneous equilibrium isentropic and zero Entropy-change processes}

Because of the ieq nature of the Brownian particle under the action of $U_0+U_1$, the particle obeys Boltzmann statistics with the instantaneous free energy $F(\lambda(t),\beta(t)$). The instantaneous entropy of the system is given by $S(t)=\beta^2(t) \phi(\lambda(t),\beta(t))$, where $\phi(\lambda,\beta)\equiv \frac{\partial F}{\partial \beta}$. Since we  have the time-dependent protocols for $\lambda(t)$ and $\beta(t)$, for given $\beta(t)$ one can adjust
$\lambda(t)$ (and vice versa) to achieve $S$ to be a constant independent of time, i.e., to achieve an isentropic and ieq process. $\lambda(t)$ can be determined simply by solving $\beta^2(t) \phi(\lambda(t),\beta(t))=$constant$=\beta_0^2 \phi(\lambda(0),\beta_0)$. 
Such a process has the unique properties of ieq and  no entropy change, but can be achieved in a finite (short) duration $\tau$. This is in marked distinction from the usual isentropic scenario  that can only be achieved in the reversible process under the quasi-static limit(infinitely slow).

\begin{figure}
\subfigure[]{ \includegraphics[width=3.in]{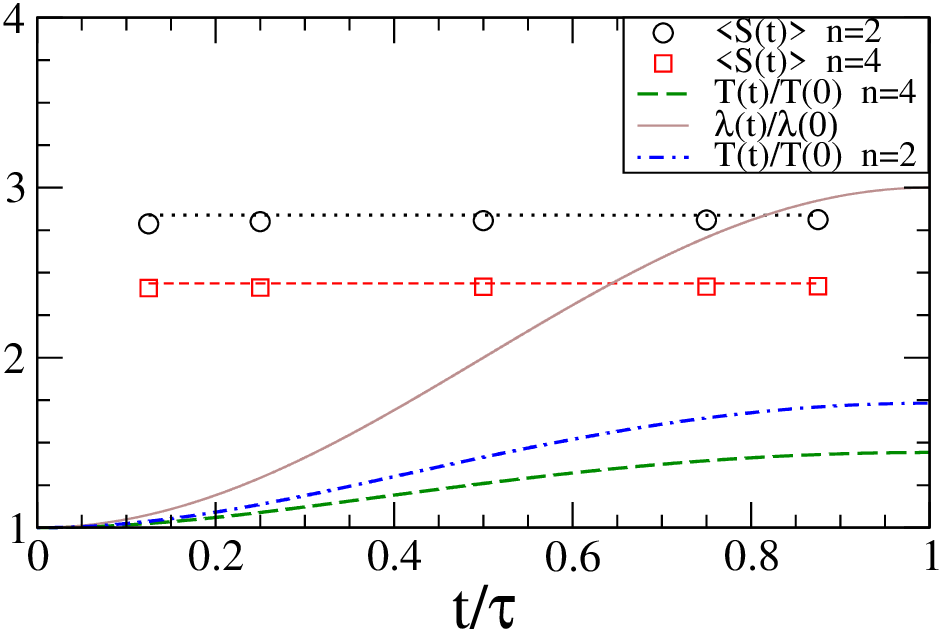}}
\subfigure[]{ \includegraphics[width=3.in]{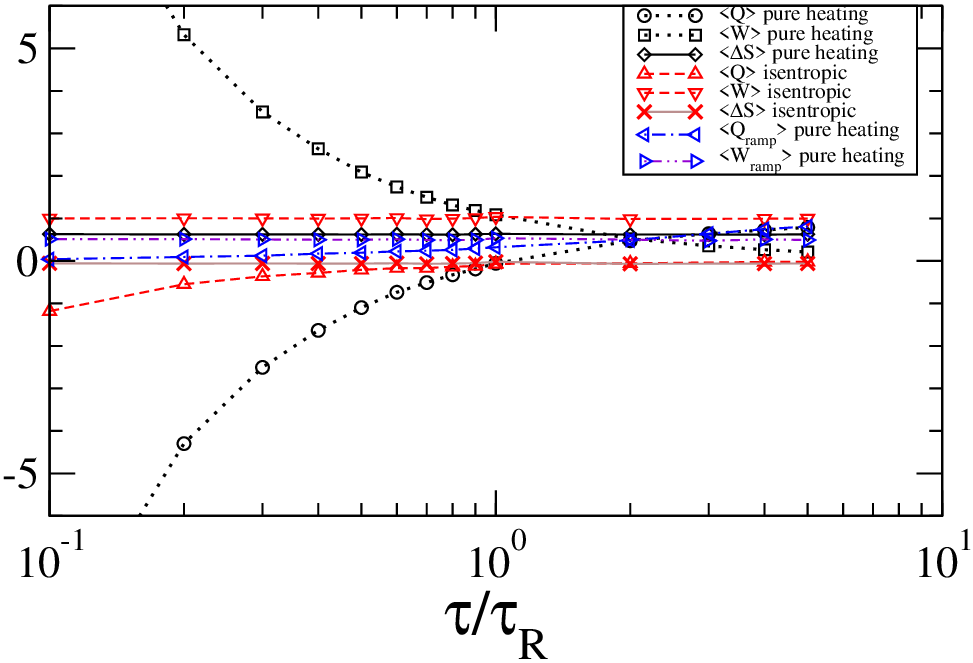}}
%\subfigure[]{ \includegraphics[width=3.in]{delSQWT1_2taum_1b.eps}}
\subfigure[]{ \includegraphics[width=3.in]{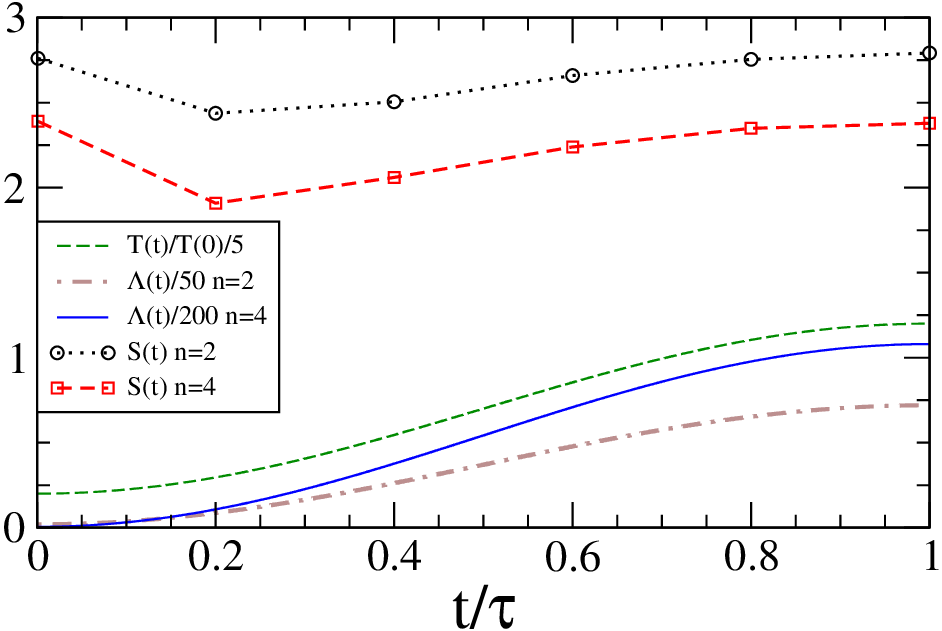}}
\subfigure[]{ \includegraphics[width=3.in]{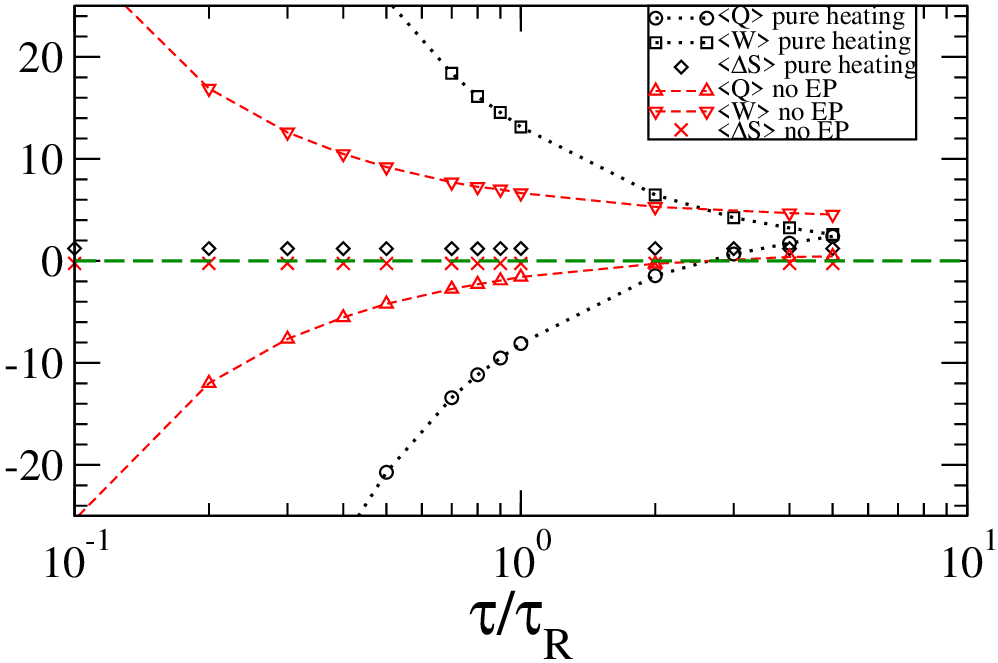}}
\caption{(a) Mean instantaneous entropy  of an underdamped Brownian particle under ieq and isentropic protocols for the  potential $U_0(x,\lambda(t))=\frac{1}{2}\lambda(t)x^n$ with $\lambda(t)=\lambda(0)+\frac{\Delta \lambda}{2}(1-\cos\frac{\pi t}{\tau})$, and $T(t)/T(0)=[\lambda(t)/\lambda(0)]^{\frac{2}{n+2}}$. $\lambda(t)/\lambda(0)=3$. The corresponding protocols for $T(t)$ and $\lambda(t)$ are also shown (curves). (b) Mean heat, work, and entropy changes plotted as a function of  the ieq transition period $\tau$. The underdamped Brownian is heated up with a temperature increase by a factor of two under ieq processes. For the ieq and isentropic process, the stiffness is varied according to (\ref{isen}). Results for the case of pure heating with the same temperature protocol are also shown for comparison. (c) Mean instantaneous entropy  of an underdamped Brownian particle under ieq and  protocols designed to have zero entropy change but $\dot{S}\neq 0$. $T(t)/T(0)=1+ \frac{1}{2}\left(\frac{T(\tau)}{T(0)}-1\right)(1-\cos\frac{\pi t}{\tau})$ and  $\Lambda(t)=1+\frac{1}{2}\left([\frac{T(\tau)}{T(0)}]^{\frac{n+2}{2}}-1\right)(1-\cos\frac{\pi t}{\tau})$. $\frac{T(\tau)}{T(0)}=6$. The instantaneous entropy $S(t)$ changes with $t$ during the ieq transition. The corresponding scaled protocols for $T(t)$ and $\Lambda(t)$ are also shown (curves). (d)  Mean heat, work, and entropy change plotted as a function of  the ieq transition period $\tau$ for the case in (c). The entropy change for the ieq process is zero (horizontal dashed line) within the uncertainties. The  mean heat, work, and entropy change for the corresponding Ramp process are also shown for comparison.}
\label{Stisen}
\end{figure}

For $U_0={1\over 2}\lambda(t) x^n$ considered here,  one can easily derive the condition for the isentropic ieq process using (\ref{St}) to give:
\begin{equation}
\lambda(t)[\beta(t)]^{1+\frac{n}{2}}=\hbox{constant}\label{isen}
\end{equation}
with the corresponding auxiliary potential is then given by 
\begin{equation}
\beta_0U_1=\frac{1}{2(n+2)}\frac{\dot{\Lambda}(t)}{\Lambda(t)}\left(x^2-2\alpha xp\right).
\end{equation}
The momentum variance is universally given by (\ref{meanx2b}) and the position  variance can be calculated to give
\begin{equation}
\alpha\langle {p_{\rm isen}}^2(t)\rangle=\frac{T(t)}{T(0)}=[\Lambda(t)]^\frac{2}{n+2}, \langle {x_{\rm isen}}^2(t)\rangle =\frac{2^\frac{2}{n}\Gamma({3\over n})}{\Gamma({1\over n})} \frac{{\beta}(t)}{\beta_0}=\tfrac{\Gamma({3\over n})}{\Gamma({1\over n})}\frac{2^\frac{2}{n}}{[\Lambda(t)]^\frac{2}{n+2}}.
\end{equation}
The corresponding work for the ieq isentropic process is
\begin{equation}
\beta_0\langle W_{\rm isen}\rangle=\tfrac{n+2}{2n}\Delta T
 +\tfrac{2^\frac{2}{n}}{n+2}
  \tfrac{\Gamma({3\over n})}{\Gamma({1\over n})}\int_0^\tau 
  \frac{\dot{\Lambda}^2}{\Lambda^{\frac{2(n+3)}{n+2}}} dt.
\end{equation}
The associated heat can be calculated using the First Law $\langle Q_{\rm isen}\rangle=\langle \Delta E\rangle -\langle W_{\rm isen}\rangle$ together with (\ref{meanE}) to give
\begin{equation}
\beta_0\langle Q_{\rm isen}\rangle=-\frac{n+2}{\beta_02^{2(1-\frac{1}{n})}}\frac{\Gamma({3\over n})}{\Gamma({1\over n})}\int_0^\tau 
  \frac{\dot{\beta}^2(t)}{\beta(t)} dt.
\end{equation} 
Contrary to the case of the quasi-static process, the ieq isentropic process is not adiabatic with $\langle Q_{\rm isen}\rangle < 0$, i.e., heat always flows out  (dissipated) of the system.
Fig. \ref{Stisen}a shows the instantaneous Shannon (Gibbs) entropy, $S(t)\equiv -\int dx dp \rho_{\rm ieq}(x,p,t) \ln \rho_{\rm ieq}(x,p,t)$, during the ieq transition under the isentropic conditions (\ref{isen}) measured from the stochastic trajectories for harmonic and non-harmonic $U_0$. 
It clearly shows that the entropy indeed is maintained at a constant value. The  heat, work, and total entropy change for the ieq isentropic transition with a temperature increased by a factor of 2 are shown in Fig. \ref{Stisen}b, together with the results for the Ramp and ieq of only the same temperature increase (pure heating with $\lambda$ fixed). 
For fast transition rates, the magnitude of the heat flowing out  under the isentropic ieq condition is much less than that of the ieq pure heating transition, indicating that  isentropic ieq transition is also a good choice for low heat dissipation.

Furthermore, one can construct an ieq process with zero entropy change but a time-vary instantaneous $S(t)$, i.e., $\Delta S=0$ and $\dot{S}\neq 0$. The protocol of $\lambda(t)$ can be arbitrary, using (\ref{DeltaS}) one just need to impose the condition for the final value of $\lambda(\tau)$ as
\begin{equation}
    \Lambda(\tau)=\left[\frac{T(\tau)}{T(0)}\right]^{\frac{n+2}{2}}.\label{EP0}
\end{equation}
Fig. \ref{Stisen}c shows the simulation results of the mean instantaneous entropy  of such a protocol with the time-varying heating  protocol (\ref{Tpropto}) and harmonic potential protocol (\ref{lamproto}), but with the final value of $\lambda$ chosen according to (\ref{EP0}). As anticipated, $\Delta S(0)=S(\tau)$ but $S(t)$ varies during the transition resulting in a zero entropy change process. The mean heat, work, and entropy change for the zero entropy change ieq process are plotted as a function of  the  transition period $\tau$ in Fig. \ref{Stisen}d, manifesting that indeed $\Delta S$ vanishes within the uncertainties. The  mean heat, work, and entropy change for the corresponding Ramp process with pure heating of the same protocol are also shown for comparison. The entropy change of the system for the Ramp process is always positive, and the associated mean heat dissipated and mean work are considerably larger than that of the zero entropy change ieq process.

\section{Time-reversed protocols: Work and Heat Relations for ieq processes}
\begin{figure}
\subfigure[]{ \includegraphics[width=\figws]{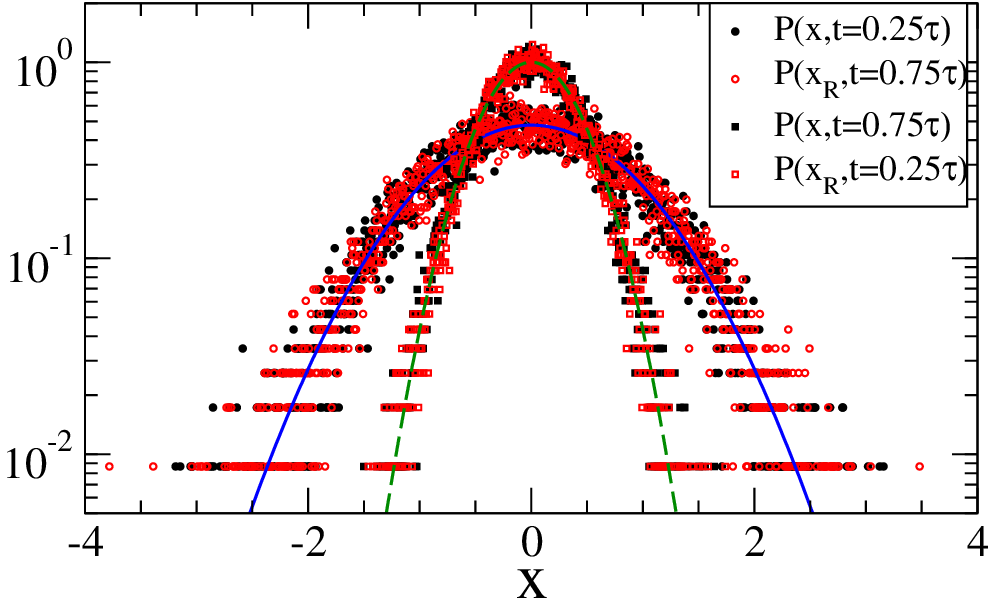}}
\subfigure[]{ \includegraphics[width=\figws]{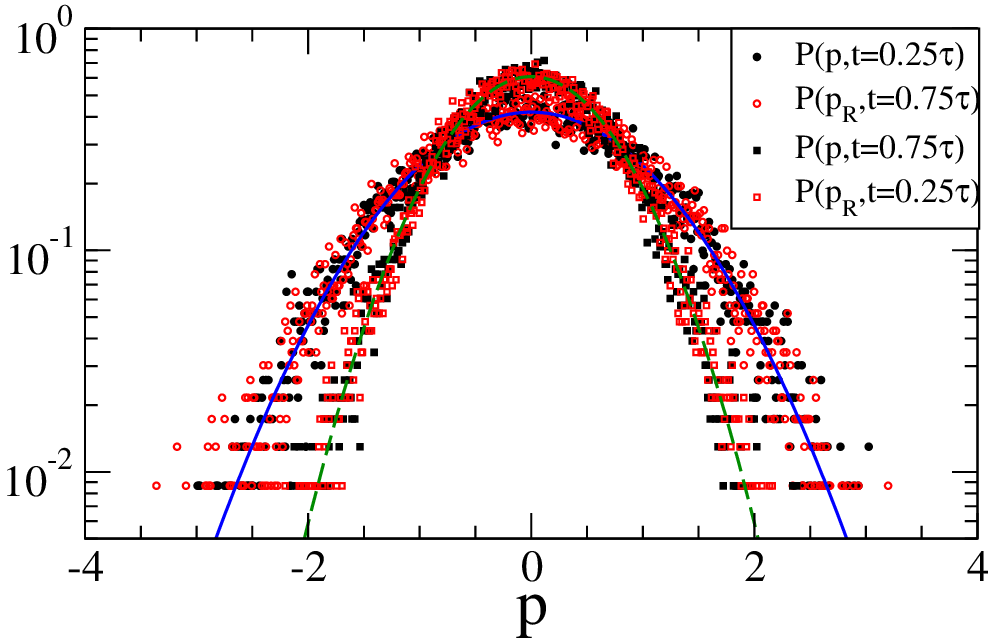}}
\subfigure[]{ \includegraphics[width=\figws]{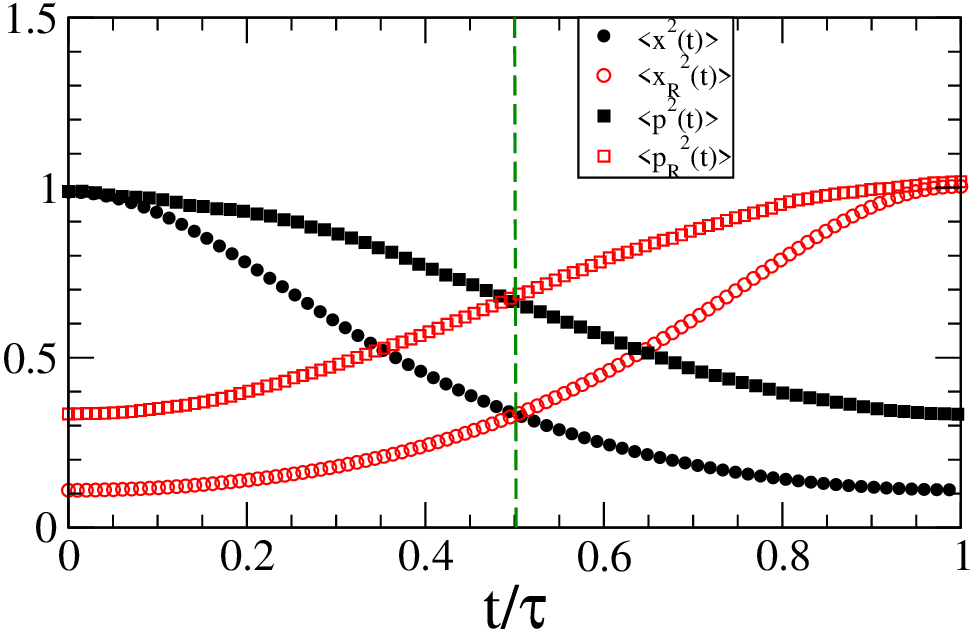}}
\subfigure[]{ \includegraphics[width=\figws]{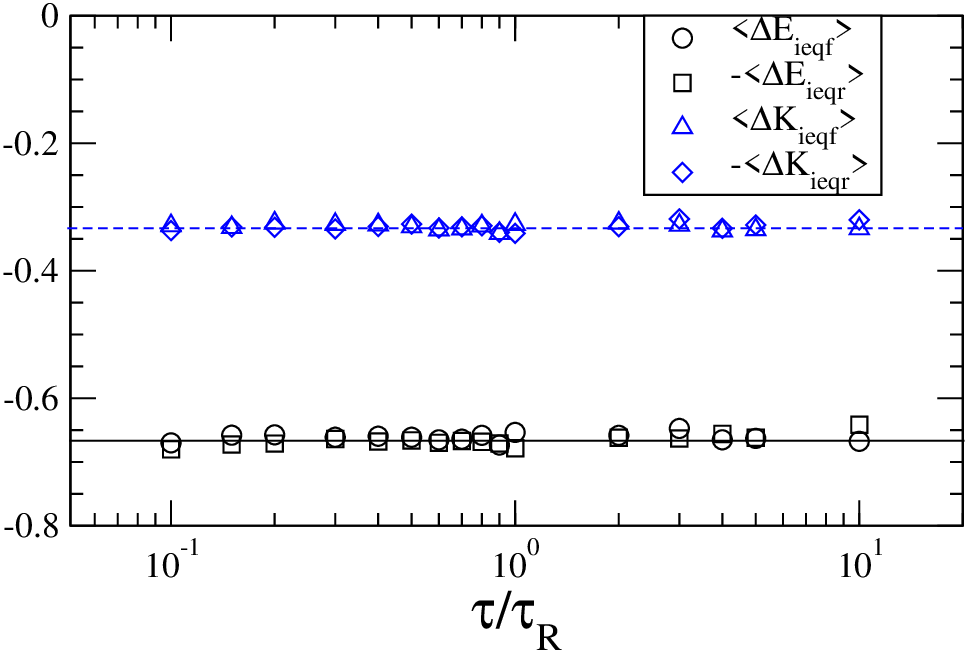}}
\caption{ (a) Position distributions of the ieqf and ieqr at times $t=0.25\tau$ and $0.75\tau$. $U_0$ is harmonic under protocols (\ref{lamproto}) and (\ref{Tpropto}) with $T(\tau)/T(0)={1\over 3}$ and $\lambda(\tau)/\lambda(0)=3$. The theoretical distributions given by (\ref{Pxt}) are also shown (curves). (b) Momentum distributions of the ieqf and ieqr as in (a). (c) Time dependence of position and momentum the variances of the ieqf and ieqr processes. The vertical dashed line marks $t=\tau/2$ is a guide to the eye for the time-reversal symmetry. (d) The mean kinetic and total energy changes for the ieqf and ieqr processes plotted as a function of $\tau$, verifying (\ref{TR}). The horizontal lines are the theoretical values given by (\ref{meanE2}).}\label{xpxpR}
\end{figure}

Here we consider  the time-reversed protocols, $\lambda_R(t)=\lambda(\tau-t)$ and $\beta_R(t)=\beta(\tau-t)$. In general, the ieq path of the reversed protocols  will be different from the time-reversed trajectory of the ieq forward process due to the odd-parity nature of $\dot{\lambda}$ and $\dot{\beta}$ in the auxiliary potential $U_1$ in (\ref{U1fg}).  
 The subscript ieqf (ieq forward) is used to differentiate it from the ieqr (ieq reverse) protocol.
The properties of the position and momentum distribution $\rho_{\rm ieq}(x,p,t)$ and the energetics of the ieq of the forward and reversed protocols are examined as follows.  
Due to the ieq nature of the position and momentum statistics, the time dependence of the distribution $\rho_{\rm ieq}(x,p,t)$ is through the instantaneous values of $\lambda(t)$ and $\beta(t)$, and hence the distribution function of the ieq paths of the time-reverse protocols is the same as the time-reversal of the ieq forward distribution function:
\begin{equation}
\rho_{\rm ieqr}(x,p,t)=\rho_{\rm ieqf}(x,p,\tau-t),\label{rhorf}
\end{equation}
i.e., the path ensembles cancel out the odd-parity in $U_1$ and restore the time-reversal symmetry of the ieq distribution.
In addition, since the $\langle E_0\rangle$, $\langle U_0\rangle$, and $S(t)$ are  state functions, one can easily see that the energy  and entropy changes of the ieqf and ieqr processes obey the following time-reversal symmetry
\begin{eqnarray}
    \langle \Delta E_{\rm ieqr}\rangle& =&-\langle \Delta E_{\rm ieqf}\rangle,\quad  \langle \Delta K_{\rm ieqr}\rangle =-\langle \Delta K_{\rm ieqf}\rangle\label{TR},\\
    \langle \Delta U_{\rm ieqr}\rangle &=&-\langle \Delta U_{\rm ieqf}\rangle,\quad\Delta S_{\rm ieqr} =-\Delta S_{\rm ieqf},\nonumber
\end{eqnarray}
which can be directly verified using (\ref{meanE}).
On the other hand, although  $Q+W=\langle \Delta E_{\rm ieq}\rangle$ which is similar to  $\langle \Delta K_{\rm ieq}\rangle+\langle \Delta U_{\rm ieq}\rangle=\langle \Delta E_{\rm ieq}\rangle$, the work and heat for the ieq process do not obey similar  time-reversal relations as in (\ref{TR}). However, remarkably one can show in the next section that the mean dissipated work and heat are the same  for ieqf and ieqr:
\begin{equation}
Q^{\rm diss}_{\rm ieqr}=Q^{\rm diss}_{\rm ieqf},\quad W^{\rm diss}_{\rm ieqr}=W^{\rm diss}_{\rm ieqf}.\label{QWdiss}
\end{equation}
Fig. \ref{xpxpR}a and  \ref{xpxpR}b respectively plot the position and momentum distributions for ieqf and ieqr processes at $t$ and $\tau-t$ for $t=\tau/4$ and $t=3\tau/4$, verifying the time-reversal symmetry of the distributions:  $P_{\rm ieqf}(x,t)=P_{\rm ieqr}(x,\tau-t)$ and $P_{G\rm ieqf}(p,t)=P_{G\rm ieqr}(p,\tau-t)$. The entire time evolution of the position and momentum variances for ieqf and ieqr are shown in Fig. \ref{xpxpR}c, clearly showing that $\langle x^2_{\rm ieqf}(t)\rangle=\langle x^2_{\rm ieqr}(\tau-t)\rangle$ and  $\langle p^2_{\rm ieqf}(t)\rangle=\langle p^2_{\rm ieqr}(\tau-t)\rangle$.
The kinetic and total energy changes for the ieqf and ieqr (times a factor of -1) processes are plotted as a function of $\tau$ in Fig. \ref{xpxpR}d, verifying that the time-reversal properties as given by (\ref{TR}) for energy changes indeed holds for ieq processes.

The ieq work and heat relations  for forward and reverse processes are related for general protocols, and  the proof of the work and heat relations is  outlined here. 
Under the ieq protocol, the Ramp potential $U_0(x,\lambda(t))$ is escorted by an auxiliary potential  given by (\ref{U1fg}) so that  the particle  experiences a total potential of  $U_{\rm ieqf}=U_0+U_1$.
Since the work rate under the ieqf protocol is $\dot {W}=\partial U_{\rm ieqf}/\partial t$, the mean work under  ieqf  is given by
\begin{eqnarray}
 \langle W_{\rm ieqf}\rangle&=&\int_0^{\tau}dt \left\langle \frac{\partial U_{\rm ieqf}}{\partial t}\right\rangle\nonumber\\
& =&\int_0^{\tau}dt \Bigl\{ \dot{\lambda}\left\langle \frac{\partial U_0}{\partial \lambda}\right\rangle+\ddot{\lambda}\langle f\rangle+   \dot{\lambda}^2 \left\langle \frac{\partial f}{\partial \lambda}\right\rangle\nonumber
  \\
& &+\ddot{\beta}\langle g\rangle+   \dot{\beta}^2 \left\langle \frac{\partial g}{\partial \beta}\right\rangle+\dot{\beta}\dot{\lambda} \left( \left\langle \frac{\partial g}{\partial \lambda}\right\rangle+  \left\langle \frac{\partial f}{\partial \beta}\right\rangle \right) \Bigr\},
\label{WF}
\end{eqnarray}
where $\langle \cdots \rangle$ is the average over the ieq Boltzmann distribution $e^{\beta(t)[F(\lambda)-H_0(x,p,\lambda(t))]}$ due to the  nature of ieq.
Under the ieq of the  reverse protocol, $\lambda_R(t)=\lambda(\tau-t)$ and $\beta_R(t)=\beta(\tau-t)$,  the particle will experience a  total potential similar to (\ref{U1fg})   as
\begin{equation}
  U_{\rm ieqr}=U_0(x,\lambda_R(t))+\dot{\lambda}_R(t) f(x,p,\lambda_R(t),\beta_R(t))+\dot{\beta}_R(t) g(x,p,\lambda_R(t),\beta_R(t)).
\end{equation}
Under the ieq of the reverse process, the work rate is ${\dot{W}}_{\rm ieqr}=\partial U_{\rm ieqr}/\partial t$. Then, the mean work is calculated in a similar way, but  the average is over the ieq Boltzmann distribution of the reverse protocol $e^{\beta_R(t)[F(\lambda_R,\beta_R)- H_0(x,\lambda_R(t))]}$ denoted by $\langle \cdots \rangle_R$:
\begin{eqnarray}
 \langle W_{\rm ieqr}\rangle_R&=&\int_0^{\tau}dt \left\langle \frac{\partial U_{{\rm ieqr}}}{\partial t}\right\rangle_R\nonumber\\
 & =&\int_0^{\tau}dt \Bigl\{ -\dot{\lambda}\left\langle \frac{\partial U_0}{\partial \lambda}\right\rangle+\ddot{\lambda}\langle f\rangle+   \dot{\lambda}^2 \left\langle \frac{\partial f}{\partial \lambda}\right\rangle\nonumber
  \\
& &+\ddot{\beta}\langle g\rangle+   \dot{\beta}^2 \left\langle \frac{\partial g}{\partial \beta}\right\rangle+\dot{\beta}\dot{\lambda} \left( \left\langle \frac{\partial g}{\partial \lambda}\right\rangle+  \left\langle \frac{\partial f}{\partial \beta}\right\rangle \right) \Bigr\}.
%  &=&\Delta F+ \int_0^{\tau}dt \left[\ddot{\lambda}\langle f\rangle+   \dot{\lambda}^2 \left\langle \frac{\partial f}{\partial \lambda}\right\rangle \right],
\label{WR}
\end{eqnarray}
Subtracting (\ref{WF}) from (\ref{WR}) gives the work relation  
\begin{equation}
     \langle W_{\rm ieqf}\rangle=\langle W_{\rm ieqr}\rangle+2\Delta \Phi, %\quad \Delta \Phi\equiv \int_0^{\tau}dt  \dot{\lambda}\left\langle \frac{\partial U_0}{\partial \lambda}\right\rangle=\Delta F- \int_0^{\tau}dt\dot{\beta}(t) \frac{ \partial F}{\partial \beta},
     \label{WWR}
\end{equation}  
where $ \Delta \Phi$ is given by (\ref{Phidef}). Since $W_{diss}\equiv \langle W\rangle - \Delta \Phi$, thus the dissipated work for ieqf and ieqr are the same. The subscript of the average of the ieq of the reverse process, $\langle \cdots \rangle_R$, can be dropped if no confusion arises.

Using the First Law of Thermodynamics, $\Delta E_{\rm ieq}=\langle Q \rangle +\langle W \rangle$, for ieqf and ieqr,  it is easy to show by similar calculations
\begin{equation}
\langle Q_{\rm ieqf}\rangle=\langle Q_{\rm ieqr}\rangle+2\int_0^{\tau}T(t)\dot{S}(t)dt, \label{QQR}
\end{equation}
i.e., the dissipated heat  (defined as $Q_{\rm diss}\equiv \langle Q\rangle-\int_0^{\tau}T(t)\dot{S}(t)dt$ ) is the same for ieqf and  ieqr processes. Furthermore, from (\ref{WWR}) and (\ref{QQR}), it is easy to see that (\ref{QWdiss}) holds.

\begin{figure}
\subfigure[]{ \includegraphics[width=3.in]{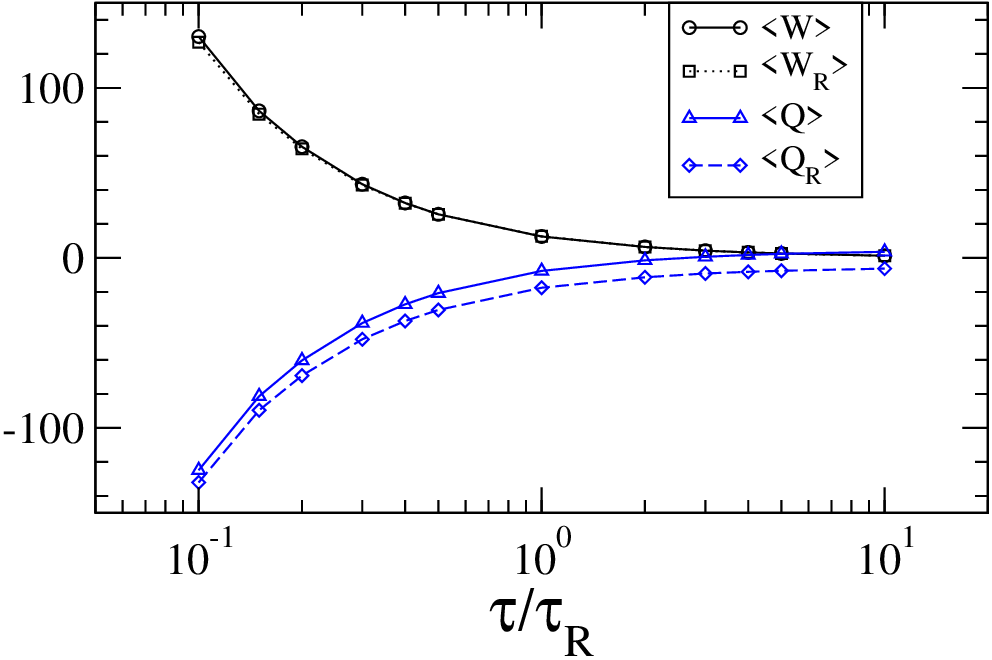}}
\subfigure[]{ \includegraphics[width=3.in]{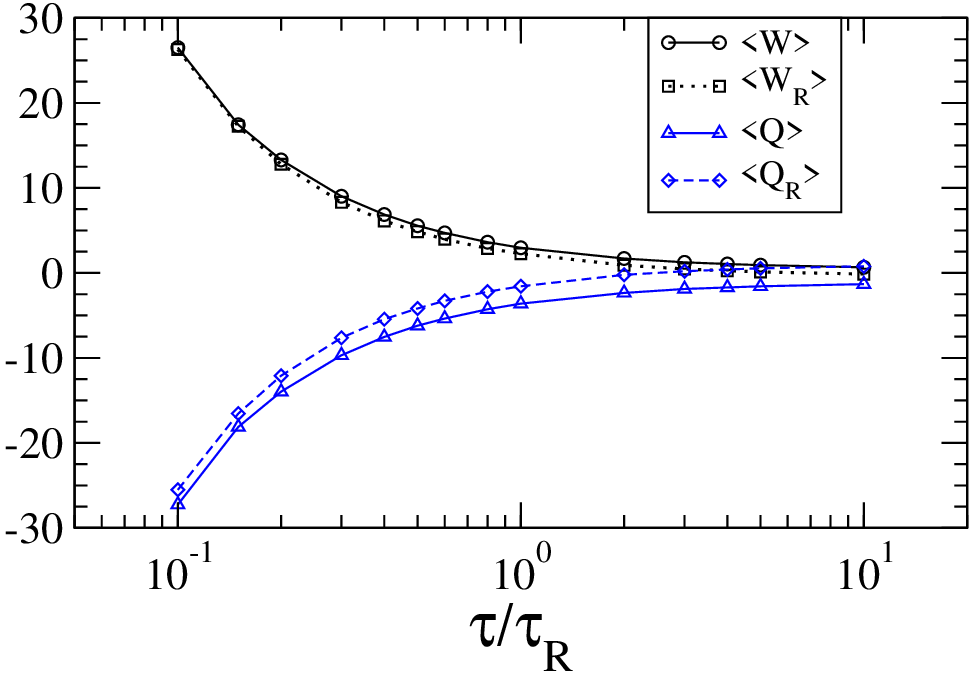}}
\subfigure[]{ \includegraphics[width=3.in]{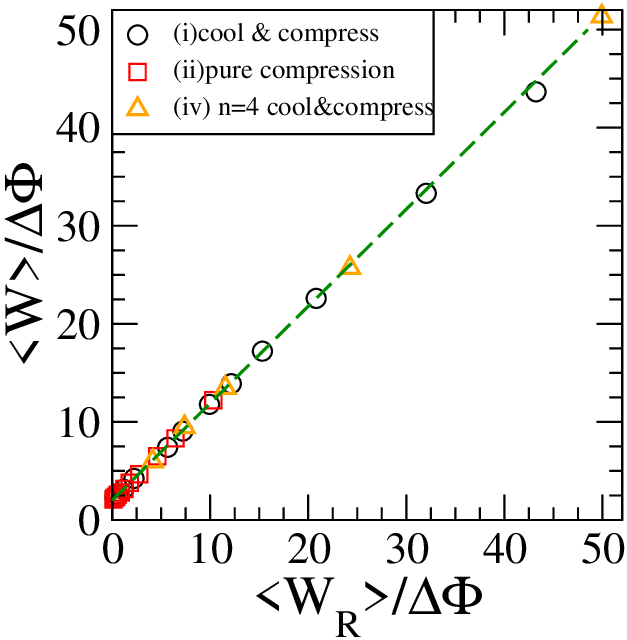}}
\subfigure[]{ \includegraphics[width=3.in]{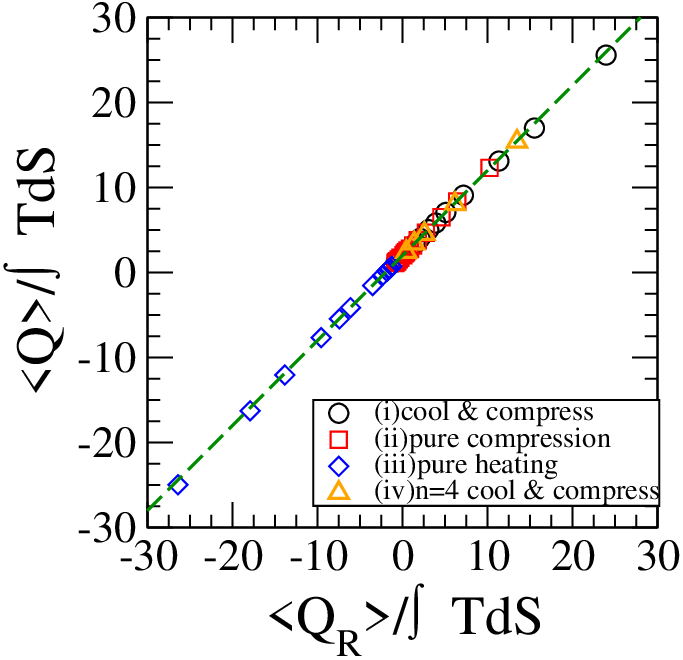}}
\caption{Work and heat in the ieq of the forward and reverse processes of a Brownian particle under harmonic potential for (a) protocols as in Fig.\ref{meanQW}a, and (b) protocols as in Fig.\ref{meanQW}b. (c) Normalized $\langle W_{\rm ieqf}\rangle$ vs. $\langle W_{\rm ieqr}\rangle$ verifying the work relation. The dashed line is the theoretical result. (i) denotes the  protocols as in Fig.\ref{meanQW}b, (ii) denotes the  protocol of pure compression of $\Lambda$ changes from 1 to 3. (iv) denotes the same protocols as (i) but for a non-harmonic ($n=4$) potential. (d)  Normalized $\langle Q{\rm ieqf}\rangle$ vs. $\langle Q_{\rm ieqr}\rangle$ verifying the heat relation. Another protocol (iii) (protocols as in Fig.\ref{meanQW}a) is also included. }\label{WQR}
\end{figure}

For $U_0={1\over 2}\lambda(t) x^n$, the direct calculation gives $ \Delta \Phi={1\over n}\int_0^{\tau}dt \frac{\dot{\Lambda}(t) }{\beta(t)\Lambda(t)}$ and $\int_0^{\tau}T(t)\dot{S}(t)dt =(\tfrac{1}{2}+\tfrac{1}{n})\Delta T -{1\over n}\int_0^{\tau}dt \frac{\dot{\Lambda}(t) }{\beta(t)\Lambda(t)}=\langle E_{\rm ieq}\rangle-\Delta \Phi$. Adding up (\ref{WWR}) and (\ref{QQR}) gives $\langle W_{\rm ieqf}\rangle+\langle Q_{\rm ieqf}\rangle-\langle E_{\rm ieqf}\rangle=\langle W_{\rm ieqr}\rangle+\langle Q_{\rm ieqr}\rangle-\langle E_{\rm ieqr}\rangle$, which is consistent with the First Law of Thermodynamics for the ieqf and ieqr processes. 
Fig. \ref{WQR}a and \ref{WQR}b show that the mean work and heat of the ieqf and ieqr processes do not obey time-reversal symmetry, contrary to the energy and entropy changes. For the case of pure heating (Fig. \ref{WQR}a) the mean works for ieqf and ieqr are the same, but are different if $\dot{\lambda}\neq 0$ (Fig. \ref{WQR}b). Moreover, the mean heats for ieqf and ieqr are different if there is a temperature-changing protocol. 
Fig. \ref{WQR}c plots the normalized ieqf mean work against that of ieqr for two different protocols for harmonic and non-harmonic potentials, showing that the work relation (\ref{WWR}) (dashed straight line) is well satisfied. The heat relation (\ref{QQR}) is illustrated to hold perfectly in Fig. \ref{WQR}d similarly.

\section{Conclusion and Outlook}
We have theoretically derived and numerically confirmed the instantaneous equilibrium transition under arbitrary time-dependent temperature and potential variations for an underdamped Brownian particle. 
The ieq protocols allow theoretical derivations of various physical quantities of interest relatively easily, and can manipulate $\lambda(t)$ and $\beta(t)$ for special purposes, such as the ieq protocols of isentropic and zero entropy change processes. Such a process with no (net) entropy change  can be viewed as a reversible process but can be carried out at a finite rate (not quasi-static).
We found that the ieq isentropic process is not adiabatic with $\langle Q_{\rm isen}\rangle<0$, i.e., heat always flows out (dissipated) of the system contrary to that of the quasi-static process.
We also derived the work relation and heat relation  for the ieq process, proving that the mean dissipated work and heat for the ieq protocols of the forward process are the same as those of the reverse process.

The instantaneous entropy $S(t)$ defined in Sec. III and the entropy change calculated in Sec. IV are the entropy of the system, i.e., the Brownian particle under ieq. To achieve ieq, the system is coupled to the external driving via the auxiliary potential which has both energy and entropy costs. Of course, the total entropy production (entropy change of the system + environment) is positive. Indeed, under ieq and isentropic conditions, the system has no change in its entropy and behaves as if it is reversible. But the system is only ‘endo-reversible’ and is coupled dissipatively to the external driving and environment. 

 In deriving the auxiliary potential $U_1$ to achieve ieq, the stochastic force acting on the underdamped Brownian particle is a Gaussian white noise whose variance is taken to be $2\gamma T(t)$, (see Appendix A) i.e. the Einstein relation  is assumed to hold instantaneously and thus one expects Fluctuation-dissipation theorem to hold instantaneously under ieq. This also is consistent with the instantaneous equi-partition result derived in (\ref{meanKt}) which holds universally. Of course, it will be instructive to calculate the correlation functions or instantaneous diffusion coefficient explicitly and also derive the response function under ieq, which will be carried in our future work. 

One possible application of the ieq process of the underdamped Brownian particle is the construction of a finite-rate colloidal heat engine, which consists of the ieq protocols of two isothermal plus  heating and cooling processes.
Since fluctuations in micro-scales cannot be ignored, a deep understanding of their physical properties is vital to the working principles and efficiencies of these micro-engines. 
To realize the finite-time heat engine, independent control of temperature and potential is essential for constructing transition paths in engine cycles. 
The ieq of the isothermal process has already been implemented \cite{Albay2020, Albay2020a} in colloid experiments while realizing  ieq protocols involving heating and cooling is challenging.
With the protocols of time-dependent potential and temperature considered in this paper, it is possible to realize an ieq of isentropic or zero entropy change process.
However, we anticipate several experimental barriers to be overcome as one tries to realize the heating and cooling protocols in an underdamped colloidal system. For example, the time scale related to inertial force is about $10^{-3}\sim 10^{-4}$ smaller than that of the position acquisition time in the experiment, indicating that the colloidal system is the overdamped one.
On the other hand, the RLC electric circuit system (under Johnson-Nyquist noise) has the same form of Langevin equations analogous to confined Brownian particles, with the charge on the capacitor corresponds to the position variable\cite{chiang2017,chiang2017b}.
Experiments with the driven RLC electric circuit influenced by thermal noise should be more feasible\cite{Freitas2020,Chang2021} because one can adjust the time scale related to inertial force in the circuit to be longer than the measurement time, resulting in an underdamped system, thanks to the wide range of the resistance of the resistors and the inductance of the inductors.
Nevertheless, achieving these experimental goals successfully will open up the broad avenue of designing microscopic heat engine cycles that can run in a much faster cycle time due to the nature of the ieq processes.
In these cycles, the distribution function of the system is well determined at any time, which might overcome the shortage of traditional models of finite-rate heat engines.

\section*{Appendix A: Derivation of the auxiliary potential $U_1(x,p,t)$}
In order to achieve instantaneous equilibrium (ieq) under the Ramp potential $U_0(x,\lambda(t))$,  a position and momentum dependent auxiliary potential $U_1(x,p,t)$ is introduced in such a way  that  the particle   experiences a total Hamiltonian 
\begin{equation}
 H=H_0(x,p,t)+ U_1(x,p,t)=\frac{p^2}{2m}+U_0(x,\lambda(t))+U_1(x,p,t),\label{U0U1udr2}
\end{equation}
and  the underdamped  Brownian particle will be at \textit{ieq} that follows the Boltzmann distribution 
\begin{equation}
\rho_{\rm ieq}(x,p,t)=e^{-\beta(t)(F(\lambda,\beta)- H_0(x,p,\lambda(t))},\label{rhoiequdr}
\end{equation}
where $ F(\lambda,\beta)=-\frac{1}{\beta(t)}\ln \int dp \int dx e^{-\beta(t) H_0(x,p,\lambda(t))}$ is the free energy of the Ramp(original) system at ieq for some instantaneous value of $\lambda$ and $\beta$ during the transition process.
The canonical (Langevin) equations for general position and momentum dependent potential reads
\begin{eqnarray}
{\dot x}&=& \frac{\partial H}{\partial p}=\frac{p}{m} + \frac{\partial U_1}{\partial p}\label{xpLang1}\\
{\dot p}&=& -\frac{\partial H}{\partial x}-\gamma \dot{x}+\xi(t)=-\frac{\partial U_0}{\partial x}-\frac{\partial U_1}{\partial x}-\gamma\left( \frac{p}{m} + \frac{\partial U_1}{\partial p}\right) +\xi(t),\label{xpLang2}
\end{eqnarray}
where  derivative with respect to $t$  is denoted by $\dot{ }$, and $\xi(t)$ is the random (Gaussian white) force with $\langle \xi(t)\rangle=0$ and  $\langle \xi(t)\xi(t')\rangle=\frac{2\gamma}{\beta(t)}\delta(t-t')$

$U_1$ can be  determined for a given Ramp potential $U_0(x,\lambda(t))$ with the protocols $\lambda(t)$  and $\beta(t)$. 
Here we shall derive the equation to determine $U_1$ for any given time-dependent potential parameter and inverse temperature protocols $\lambda(t)$ and $\beta(t)$. And for the Ramp potential of the form  $U_0(x,\lambda(t))={1\over 2}\lambda(t)x^n$, expression for $U_1$ is derived explicitly.

$U_1$ is determined by the requirement that  $\rho_{\rm ieq}(x,t)$ must satisfy the Kramers equation
\begin{equation}
\frac{\partial  \rho_{\rm ieq}}{\partial t}=-\frac{\partial }{\partial x}\left[ \rho_{\rm ieq}\left( \frac{p}{m}+\frac{\partial  U_1}{\partial p}\right)\right] + \frac{\partial }{\partial p}\left[ \rho_{\rm ieq}\left(  \frac{\partial U_0}{\partial x}+\frac{\partial U_1}{\partial x}+ \frac{\gamma p}{m} + \gamma\frac{\partial U_1}{\partial p}  +\frac{\gamma }{\beta(t)\rho_{\rm ieq}}\frac{\partial  \rho_{\rm ieq}}{\partial p}  \right)\right].\label{kramers}
\end{equation}
Upon substituting (\ref{rhoiequdr}) into (\ref{kramers}),   one obtains
\begin{equation}
\frac{\gamma }{\beta} \frac{\partial^2  U_1}{\partial p^2}+\left(\frac{\partial U_0}{\partial x}-\frac{\gamma p}{m} \right)\frac{\partial U_1}{\partial p}-  \frac{ p}{m} \frac{\partial U_1}{\partial x}=\frac{\dot{\beta}}{\beta}(F-H_0)+\dot{\beta}\frac{\partial F}{\partial \beta}+\left( \frac{\partial F}{\partial \lambda}- \frac{\partial U_0}{\partial \lambda}\right)\dot{\lambda}\label{U1PDEapp}
\end{equation}
which is a linear PDE  for $U_1(x,p,t)$. Because of the linear dependence of
$\dot{\lambda}$ and $\dot{\beta}$ in the RHS of (\ref{U1PDEapp}), it is easy to see that $U_1$ is of the form given by (\ref{U1fg}) with the $f$ and $g$ determined from the de-coupled linear PDEs
\begin{eqnarray}
\frac{\gamma }{\beta} \frac{\partial^2  f}{\partial p^2}+\left(\frac{\partial U_0}{\partial x}-\frac{\gamma p}{m} \right)\frac{\partial f}{\partial p}-  \frac{ p}{m} \frac{\partial f}{\partial x}= \frac{\partial F}{\partial \lambda}- \frac{\partial U_0}{\partial \lambda} \label{fPDE}\\
\frac{\gamma }{\beta} \frac{\partial^2  g}{\partial p^2}+\left(\frac{\partial U_0}{\partial x}-\frac{\gamma p}{m} \right)\frac{\partial g}{\partial p}-  \frac{ p}{m} \frac{\partial g}{\partial x}=\frac{1}{\beta}(F-H_0)+\frac{\partial F}{\partial \beta},\label{gPDE}
\end{eqnarray}
which can be solved separately (say using the power series expansion method).
We now solve  $f$ and $g$ for the case of $U_0(x,\lambda)={1\over 2} \lambda x^n$. In this case $\frac{\partial U_0}{\partial x}={n\over 2}\lambda x^{n-1}$, $F(\lambda,\beta)$ is given by (\ref{Flambet}) and hence $\frac{\partial F}{\partial \lambda}=\frac{1}{n\beta\lambda}$, $\frac{\partial F}{\partial \beta}=\frac{1}{\beta^2}({1\over 2}+{1\over n})-\frac{F}{\beta}$. Thus the RHS of (\ref{fPDE}) and (\ref{gPDE}) are $\frac{1}{n\beta\lambda}-{1\over 2} x^n$ and $\frac{1}{\beta^2}({1\over 2}+{1\over n})-\frac{p^2}{2m\beta}-\frac{\lambda x^n}{2m\beta}$ respectively.
$g$ can be solved by the trial form of $g=a_1 p^2+a_2xp +a_3 x^2 +a_4 x^n$  ($a$'s are constants). Upon substitution into (\ref{gPDE})and matching the coefficients of $p^2$, $xp$,$x^n$ and $x^{n-1} p$, one gets $a_1=\frac{1}{\gamma\beta}({1\over 2}+{1\over n})$, $a_2=-\frac{1}{n\beta}$, $a_3=\frac{\gamma}{2n\beta}$, $a_4=\frac{m\lambda}{2 \gamma\beta}({1\over 2}+{1\over n})$. $f$ can be determined in a similar way, summarizing, the solutions for $f$ and $g$ are
\begin{eqnarray}
f&=& \frac{m}{2n\gamma\lambda}\left[\frac{({ p}-\gamma{ x} )^2}{m}+ \lambda{ x}^n  \right]  \\
g&=&  \frac{1}{2\gamma\beta}({1\over 2}+{1\over n})(p^2+m\lambda x^n)-\frac{1}{n\beta}x(p-\frac{\gamma}{2}x).
\end{eqnarray}
Then $U_1=\dot{\lambda}f(x,p,\lambda)+ \dot{\beta}g(x,p,\lambda,\beta)$ which can be expressed as (\ref{U1udrxn}) or (\ref{U1xpxn2}) .
 Notice that $U_1=\dot{\lambda}f(x,p,\lambda)$ is the auxiliary potential for an underdamped Brownian particle at the fixed temperature, which agrees with the result in \cite{Li2017}.

\section*{Appendix B: Simulation details of the underdamped Langevin systems} 
In terms of the dimensionless position and momentum, 
 with $U_0$ and $U_1$ given by (\ref{U0xn2}) and 
  (\ref{U1xpxn2}), the direct calculation gives
  \begin{eqnarray}
   \beta_0\tfrac{\partial U_1}{\partial p}&=&\tfrac{\alpha}{n}\nu(t)(\alpha p-x)+\tfrac{\alpha^2}{2}\tfrac{\dot{\beta}(t)}{\beta(t)} p\\
 \beta_0\tfrac{\partial (U_0+U_1)}{\partial x}&=&[  \tfrac{n}{2}+ \tfrac{n\alpha}{4}\tfrac{\dot{\beta}(t)}{\beta(t)}+\tfrac{\alpha}{2}\nu(t)]\Lambda(t)x^{n-1}-\tfrac{\nu(t)}{n}(\alpha p-x).
\end{eqnarray}
From the canonical equations of motion (\ref{xpLang1}) and (\ref{xpLang2}), the Langevin equations of the ieq process under $U_0$ of the form (\ref{U0xn}), read
\begin{eqnarray}
\frac{dx}{dt}&=&p+\left[\frac{\nu(t)}{n}+\frac{\dot{\beta}(t)}{2\beta_0} \right]\alpha p-\frac{\nu(t)}{n}x \label{xLang}\\
\frac{dp}{dt}&=&-\left[{1\over\alpha}+\frac{\dot{\beta}}{2\beta_0} \right]p -\left[\frac{\nu(t)}{2}+\frac{n\dot{\beta}(t)}{4\beta_0} +\frac{n}{2\alpha}\right]\Lambda(t) x^{n-1} + \xi(t)\label{pLang}\\
 & &\langle\xi(t)\xi(t')\rangle=\frac{2\beta_0}{\alpha^2\beta(t)}\delta(t-t').
\end{eqnarray}
 
 According to the original idea of Langevin\cite{Langevin}, heat transfer is due to the collision around the Brownian particle by the viscous and random forces given by $-\gamma \dot{x}+ \xi(t)$ and the stochastic heat is given by $dQ=(-\gamma \dot{x}+ \xi(t))\circ dx$, where $\circ$ denotes the Stratonovich notation. Then using (\ref{xpLang2}), one gets the ieq stochastic heat $dQ=(\frac{p}{m}+\partial_pU_1)\circ dp +\partial_x(U_0+U_1) \circ dx$, or in terms of the dimensionless variables, the ieq heat is given by
   \begin{equation}
 \beta_0dQ= \tfrac{\alpha}{2}dp^2 +\beta_0\tfrac{\partial (U_0+U_1)}{\partial x}\circ dx + \beta_0\tfrac{\partial U_1}{\partial p}\circ dp.\label{dQapp}
  \end{equation}
  The change in total energy in an infinitesimal ieq path is simply
  \begin{equation}
  dE_{\rm ieq}=\frac{\alpha}{2\beta_0}dp^2+dU_0+dU_1
  \end{equation}
and together with (\ref{dQapp}) one gets
\begin{equation}
dE_{\rm ieq}-dQ=\frac{\partial( U_0+U_1)}{\partial t}\equiv dW
\end{equation}
verifying the First Law of thermodynamics.

The Langevin equations (\ref{xLang}) and (\ref{pLang}) are solved numerically using the Euler-Maruyama scheme with a time step of $\delta t=10^{-4}$, and ensemble averages are computed using 5000 to 20000 stochastic trajectories.

\begin{acknowledgments}  This work  has been supported by the Ministry of Science and Technology of Taiwan
 under grant no.  110-2112-M-008-026-MY3 (PYL) and 109-2112-M-008-006- (YJ), and NCTS of Taiwan.
\end{acknowledgments} 
\bibliographystyle{unsrt}
\bibliography{references}

\begin{thebibliography}{10}

\bibitem{Callen1985ThermodynamicsThermostatistics}
Herbert~B Callen.
\newblock {\em {Thermodynamics and an Introduction to Thermostatistics}}.
\newblock John wiley {\&} sons, 1985.

\bibitem{Jarzynski1997}
C~Jarzynski.
\newblock {Equilibrium free-energy differences from nonequilibrium
  measurements: A master-equation approach}.
\newblock {\em Phys. Rev. E}, 56(5):5018--5035, 11 1997.

\bibitem{Crooks1999}
G~E Crooks.
\newblock {Entropy production fluctuation theorem and the nonequilibrium work
  relation for free energy differences.}
\newblock {\em Phys. Rev. E}, 60(3):2721--6, 9 1999.

\bibitem{Sekimoto2010}
Ken Sekimoto.
\newblock {\em {Stochastic Energetics}}, volume 799 of {\em Lecture Notes in
  Physics}.
\newblock Springer Berlin Heidelberg, Berlin, Heidelberg, 2010.

\bibitem{Seifert2012}
Udo Seifert.
\newblock {Stochastic thermodynamics, fluctuation theorems and molecular
  machines.}
\newblock {\em Rep. Prog. Phys.}, 75(12):126001, 12 2012.

\bibitem{Carberry2004FluctuationsTrap}
D.~Carberry, J.~Reid, G.~Wang, E.~Sevick, Debra Searles, and Denis Evans.
\newblock {Fluctuations and Irreversibility: An Experimental Demonstration of a
  Second-Law-Like Theorem Using a Colloidal Particle Held in an Optical Trap}.
\newblock {\em Phys. Rev. Lett.}, 92(14):140601, 4 2004.

\bibitem{Blickle2011}
Valentin Blickle and Clemens Bechinger.
\newblock {Realization of a micrometre-sized stochastic heat engine}.
\newblock {\em Nat. Phys.}, 8(2):143--146, 12 2011.

\bibitem{Martinez2015}
I.~A. Mart{\'{i}}nez, {\`{E}}.~Rold{\'{a}}n, Luis Dinis, Dmitri Petrov, Juan
  M.~R. Parrondo, and R.~A. Rica.
\newblock {Brownian Carnot engine}.
\newblock {\em Nature Physics}, 12(1):67--70, 10 2015.

\bibitem{Collin2005VerificationEnergies}
D~Collin, F~Ritort, C~Jarzynski, S~B Smith, I~Tinoco, and C~Bustamante.
\newblock {Verification of the Crooks fluctuation theorem and recovery of RNA
  folding free energies.}
\newblock {\em Nature}, 437(7056):231--4, 9 2005.

\bibitem{Liphardt2002EquilibriumEquality}
Jan Liphardt, Sophie Dumont, Steven~B Smith, Ignacio Tinoco, and Carlos
  Bustamante.
\newblock {Equilibrium information from nonequilibrium measurements in an
  experimental test of Jarzynski's equality.}
\newblock {\em Science}, 296(5574):1832--5, 6 2002.

\bibitem{Batalhao2014ExperimentalSystem}
Tiago~B. Batalh{\~{a}}o, Alexandre~M. Souza, Laura Mazzola, Ruben Auccaise,
  Roberto~S. Sarthour, Ivan~S. Oliveira, John Goold, Gabriele De~Chiara, Mauro
  Paternostro, and Roberto~M. Serra.
\newblock {Experimental Reconstruction of Work Distribution and Study of
  Fluctuation Relations in a Closed Quantum System}.
\newblock {\em Phys. Rev. Lett.}, 113(14):140601, 10 2014.

\bibitem{An2015ExperimentalSystem}
Shuoming An, Jing~Ning Zhang, Mark Um, Dingshun Lv, Yao Lu, Junhua Zhang,
  Zhang~Qi Yin, H.~T. Quan, and Kihwan Kim.
\newblock {Experimental test of the quantum Jarzynski equality with a
  trapped-ion system}.
\newblock {\em Nat. Phys.}, 11(2):193--199, 2015.

\bibitem{Albay2021}
John A~C Albay, Zhi-Yi Zhou, Cheng-Hung Chang, and Yonggun Jun.
\newblock {Shift a laser beam back and forth to exchange heat and work in
  thermodynamics}.
\newblock {\em Scientific Reports}, 11(1):4394, dec 2021.

\bibitem{Rossnagel2016}
J.~Rossnagel, Samuel~T Dawkins, Karl~N Tolazzi, Obinna Abah, Eric Lutz,
  F.~Schmidt-Kaler, and Kilian Singer.
\newblock {A single-atom heat engine}.
\newblock {\em Science}, 352(6283):325--329, 4 2016.

\bibitem{Schmiedl2008}
T.~Schmiedl and U.~Seifert.
\newblock {Efficiency at maximum power: An analytically solvable model for
  stochastic heat engines}.
\newblock {\em EPL (Europhysics Letters)}, 81(2):20003, 1 2008.

\bibitem{Esposito2010}
Massimiliano Esposito, Ryoichi Kawai, Katja Lindenberg, and Christian Van~den
  Broeck.
\newblock Efficiency at maximum power of low-dissipation carnot engines.
\newblock {\em Phys. Rev. Lett.}, 105:150603, 2010.

\bibitem{Verley2014}
Gatien Verley, Massimiliano Esposito, Tim Willaert, and Christian {Van den
  Broeck}.
\newblock {The unlikely Carnot efficiency}.
\newblock {\em Nat. Commun.}, 5(May):4721, sep 2014.

\bibitem{Polettini2015}
M.~Polettini, G.~Verley, and M.~Esposito.
\newblock {Efficiency Statistics at All Times: Carnot Limit at Finite Power}.
\newblock {\em Phys. Rev. Lett.}, 114(5):050601, feb 2015.

\bibitem{Manikandan2019}
Sreekanth~K. Manikandan, Lennart Dabelow, Ralf Eichhorn, and Supriya
  Krishnamurthy.
\newblock {Efficiency fluctuations in microscopic machines}.
\newblock {\em Phys. Rev. Lett.}, 122:140601, 2019.

\bibitem{Gingrich2016}
Todd~R. Gingrich, Jordan~M. Horowitz, Nikolay Perunov, and Jeremy~L. England.
\newblock Dissipation bounds all steady-state current fluctuations.
\newblock {\em Phys. Rev. Lett.}, 116:120601, 2016.

\bibitem{Pietzonka2018}
Patrick Pietzonka and Udo Seifert.
\newblock {Universal Trade-Off between Power, Efficiency, and Constancy in
  Steady-State Heat Engines}.
\newblock {\em Phys. Rev. Lett.}, 120(19):190602, 2018.

\bibitem{Li2019TUR}
Junang Li, Jordan~M. Horowitz, Todd~R. Gingrich, and Nikta Fakhri.
\newblock {Quantifying dissipation using fluctuating currents}.
\newblock {\em Nat. Commun.}, 10(1):1666, dec 2019.

\bibitem{Guery-Odelin2019a}
D.~Gu{\'{e}}ry-Odelin, A.~Ruschhaupt, A.~Kiely, E.~Torrontegui,
  S.~Mart{\'{i}}nez-Garaot, and J.~G. Muga.
\newblock {Shortcuts to adiabaticity: Concepts, methods, and applications}.
\newblock {\em Rev. Mod. Phys.}, 91(4):045001, 10 2019.

\bibitem{Ibanez2012}
S.~Ib{\'{a}}{\~{n}}ez, Xi~Chen, E.~Torrontegui, J.~G. Muga, and A.~Ruschhaupt.
\newblock {Multiple Schr{\"{o}}dinger Pictures and Dynamics in Shortcuts to
  Adiabaticity}.
\newblock {\em Phys. Rev. Lett.}, 109(10):100403, 9 2012.

\bibitem{Campbell2017}
Steve Campbell and Sebastian Deffner.
\newblock Trade-off between speed and cost in shortcuts to adiabaticity.
\newblock {\em Phys. Rev. Lett.}, 118:100601.

\bibitem{Patra2017ShortcutsFields}
Ayoti Patra and Christopher Jarzynski.
\newblock {Shortcuts to adiabaticity using flow fields}.
\newblock {\em New Journal of Physics}, 19(12):125009, 12 2017.

\bibitem{Martinez2016}
Ignacio~A. Mart{\'{i}}nez, Artyom Petrosyan, David Gu{\'{e}}ry-Odelin, Emmanuel
  Trizac, and Sergio Ciliberto.
\newblock {Engineered swift equilibration of a Brownian particle}.
\newblock {\em Nature Physics}, 12(9):843--846, 9 2016.

\bibitem{Chupeau2020}
Marie Chupeau, Jannes Gladrow, Alexei Chepelianskii, Ulrich~F. Keyser, and
  Emmanuel Trizac.
\newblock {Optimizing Brownian escape rates by potential shaping}.
\newblock {\em Proc. Natl. Acad. Sci. U.S.A.}, 117(3):1383--1388, 1 2020.

\bibitem{Plata2020}
Carlos~A. Plata, David Gu\'ery-Odelin, Emmanuel Trizac, and Antonio Prados.
\newblock Finite-time adiabatic processes: Derivation and speed limit.
\newblock {\em Phys. Rev. E}, 101:032129, Mar 2020.

\bibitem{cabara2020}
Yoseline Rosales-Cabara, Giovanni Manfredi, Gabriel Schnoering, Paul-Antoine
  Hervieux, Laurent Mertz, and Cyriaque Genet.
\newblock Optimal protocols and universal time-energy bound in brownian
  thermodynamics.
\newblock {\em Phys. Rev. Research}, 2:012012(R), Jan 2020.

\bibitem{Li2017}
Geng Li, H~T Quan, and Z~C Tu.
\newblock {Shortcuts to isothermality and nonequilibrium work relations}.
\newblock {\em Phys. Rev. E}, 96(1):012144, 7 2017.

\bibitem{Li2019}
Geng Li and Z.~C. Tu.
\newblock {Stochastic thermodynamics with odd controlling parameters}.
\newblock {\em Phys. Rev. E}, 100(1):012127, jul 2019.

\bibitem{Albay2019}
John A.~C. Albay, Sarah~R. Wulaningrum, Chulan Kwon, Pik-Yin Lai, and Yonggun
  Jun.
\newblock {Thermodynamic cost of a shortcuts-to-isothermal transport of a
  Brownian particle}.
\newblock {\em Phys. Rev. Res.}, 1(3):033122, 11 2019.

\bibitem{Albay2020}
John A.~C. Albay, Chulan Kwon, Pik-Yin Lai, and Yonggun Jun.
\newblock {Work relation in instantaneous-equilibrium transition of forward and
  reverse processes}.
\newblock {\em New Journal of Physics}, 22(12):123049, dec 2020.

\bibitem{Albay2020a}
John A.~C. Albay, Pik-Yin Lai, and Yonggun Jun.
\newblock {Realization of finite-rate isothermal compression and expansion
  using optical feedback trap}.
\newblock {\em Applied Physics Letters}, 116(10):103706, mar 2020.

\bibitem{chiang2017}
K.-H. Chiang, C.-L. Lee, P.-Y. Lai, and Y.-F. Chen.
\newblock Entropy production and irreversibility of dissipative trajectories in
  electric circuits.
\newblock {\em Phys. Rev. E}, 95:012158, Jan 2017.

\bibitem{chiang2017b}
K.-H. Chiang, C.-L. Lee, P.-Y. Lai, and Y.-F. Chen.
\newblock Electrical autonomous brownian gyrator.
\newblock {\em Phys. Rev. E}, 96:032123, Sep 2017.

\bibitem{Freitas2020}
Nahuel Freitas, Jean-Charles Delvenne, and Massimiliano Esposito.
\newblock {Stochastic and Quantum Thermodynamics of Driven RLC Networks}.
\newblock {\em Physical Review X}, 10(3):031005, jul 2020.

\bibitem{Chang2021}
Hsin Chang, Kuan-hsun Chiang, Yonggun Jun, Pik-yin Lai, and Yung-fu Chen.
\newblock {Generation of virtual potentials by controlled feedback in electric
  circuit systems}.
\newblock {\em Physical Review E}, 103(4):042138, apr 2021.

\bibitem{Langevin}
P.~Langevin.
\newblock {Sur la théorie de mouvement Brownien}.
\newblock {\em C. R. Hebd. Seances Acad. Sci}, 146:530, 1908.

\end{thebibliography}

\end{document}